\documentclass[iicol,pdflatex,sn-standardnature]{sn-jnl} 
\jyear{2022}%
\usepackage{dblfloatfix}    
\usepackage{multirow}

\usepackage{xcolor}
\definecolor{codegreen}{rgb}{0,0.6,0}
\definecolor{codegray}{rgb}{0.5,0.5,0.5}
\definecolor{codepurple}{rgb}{0.58,0,0.82}
\definecolor{backcolour}{rgb}{0.95,0.95,0.92}
\definecolor{BLACK}{RGB}{0,0,0}
\lstdefinestyle{mystyle}{
    backgroundcolor=\color{backcolour},   
    commentstyle=\color{codegreen},
    keywordstyle=\color{magenta},
    numberstyle=\tiny\color{codegray},
    stringstyle=\color{codepurple},
    basicstyle=\ttfamily\footnotesize,
    breakatwhitespace=false,         
    breaklines=true,                 
    captionpos=b,                    
    keepspaces=true,                 
    numbers=left,                    
    numbersep=5pt,                  
    showspaces=false,                
    showstringspaces=false,
    showtabs=false,                  
    tabsize=2
}
\usepackage{listings}

\colorlet{SB}{purple!50!black}   
\colorlet{DS}{blue!50!black}     
\colorlet{FB}{green!50!black}    

\usepackage[normalem]{ulem}
\usepackage{siunitx}

\begin{document}

\title[Article Title]{IBISCape: A Simulated Benchmark for multi-modal SLAM Systems Evaluation in Large-scale Dynamic Environments}

\author*[1]{\fnm{Abanob} \sur{Soliman}}\email{abanob.soliman@univ-evry.fr}

\author[1]{\fnm{Fabien} \sur{Bonardi}}\email{fabien.bonardi@ibisc.univ-evry.fr}

\author[1]{\fnm{D\'esir\'e} \sur{Sidib\'e}}\email{drodesire.sidibie@ibisc.univ-evry.fr}

\author[1]{\fnm{Samia} \sur{Bouchafa}}\email{samia.bouchafa@ibisc.univ-evry.fr}

\affil*[1]{\orgdiv{Universit\'e Paris-Saclay, Univ Evry}, \orgname{IBISC Laboratory}, \orgaddress{\street{34 Rue du Pelvoux}, \city{Evry}, \postcode{91020}, \state{Essonne}, \country{France}}}

\abstract{The development process of high-fidelity SLAM systems depends on their validation upon reliable datasets. Towards this goal, we propose IBISCape, a simulated benchmark that includes data synchronization and acquisition APIs for telemetry from heterogeneous sensors: stereo-RGB/DVS, Depth, IMU, and GPS, along with the ground truth scene segmentation and vehicle ego-motion. Our benchmark is built upon the CARLA simulator, whose back-end is the Unreal Engine rendering a high dynamic scenery simulating the real world. Moreover, we offer 34 multi-modal datasets suitable for autonomous vehicles navigation, including scenarios for scene understanding evaluation like accidents, along with a wide range of frame quality based on a dynamic weather simulation class integrated with our APIs. We also introduce the first calibration targets to CARLA maps to solve the unknown distortion parameters problem of CARLA simulated DVS and RGB cameras. Finally, using IBISCape sequences, we evaluate four ORB-SLAM3 systems (monocular RGB, stereo RGB, Stereo Visual Inertial (SVI), and RGB-D) performance and BASALT Visual-Inertial Odometry (VIO) system on various sequences collected in simulated large-scale dynamic environments.}


\maketitle

\section{Introduction}

Autonomous vehicles navigating in unknown and dynamic environments need to rely on accurate perception systems for real-time 3D mapping. These perception systems must function optimally in all weather conditions and situations. That enables the vehicle to make decisions for its passengers or the surrounding pedestrians and cars. To this objective, many novel technologies have been developed over the last decade. Some use vision sensors such as monocular Visual Odometry (VO) \cite{7782863}, which can suffer from estimations up to a scale factor. Innovative solutions to estimate this scale factor by fusion with another sensor like mono/stereo Visual-Inertial Odometry (VIO) \cite{Leutenegger2014,7557075,8421746} and RGB-D SLAM \cite{Kerl2013} to add depth information have been proposed. Other works use LiDAR \cite{Alliez2020RealTimeMS} sensor that provides high precision point clouds mapping of the scene, or use the GPS \cite{caron2006gps} for localization using satellite signal triangulation.

Multi-modal datasets can enrich and broaden the research in the Simultaneous Localization and Mapping (SLAM) field, mainly applied to autonomous ground vehicle (AGV) navigation in large-scale and dynamic environments. These environments have specific characteristics, such as the dynamic range of the objects' intensities in the scene. For example: mapping a small room with proper lighting can be of higher quality than mapping a road in a city (large-scale) at night with high-intensity fog, rain, and wind (dynamic environment). The advantages of system multi-modality appear when depending on cameras with high dynamic range, such as the DAVIS sensor and regular low-cost cameras and sensors (IMU/GPS). This multi-modality leads to completing the data shortages during the scene mapping and AGVs localization. 

Nowadays, multi-modal frameworks of sensors have proven to be attracting the attention of many researchers in robotics perception for different tasks such as calibration \cite{yangicalib, 9387269} and odometry \cite{leeefficient, gehrig2021combining}. That is due to the fact that heterogeneous sensors that perceive the environment allow the acquisition of complementary information data about the scene. Moreover, sensors multi-modality can also include redundancy such as stereo-DVS or stereo-RGB cameras configurations. Having redundancy in the system sensors can improve both the precision and the quality of the collected scene landmarks. Furthermore, some sensors have a high temporal resolution and are sensitive to the scene intensity changes, such as the DAVIS sensor (Event Camera) \cite{Gehrig21ral}. While other sensors can efficiently detect and track landmarks and scene features in the 3D spatial domain, such as RGB-D cameras \cite{Li2021PlanarSLAM} and LiDAR \cite{s20072068}. 

Simulated datasets provide the possibility to have sequences in various complex scenarios. Moreover, setting a hardware data acquisition framework with a specific configuration can be costly and time-consuming and is prone to multiple limitations such as the carrier (car, handheld, drone), weather conditions, sensors configuration, and synchronization. Furthermore, open sourcing the data acquisition APIs with configurable calibration targets can widen the research horizon in multi-modal calibration and sensors synchronization to reach reliable and easy algorithms to implement.

\textbf{Our main contributions} to mitigate all these hardware configuration constraints and to facilitate the multi-modal data synchronization and acquisition process are:
\begin{itemize}
\item The IBISCape benchmark of 34 sequences for multi-modal VI-SLAM applications, besides open-sourcing our multi-modal data acquisition APIs.
\item A simulated core sensor suite of most visual-inertial sensors used in assessing visual SLAM systems, along with providing high-resolution frames of variable quality depending on the dynamic level of the scene. The full sensor setup is represented in Fig. \ref{fig:IBISCape}. 
\item A solution to calibrate CARLA \cite{Dosovitskiy17} RGB and DVS cameras with unknown distortion values by introducing the first high quality calibration targets to one of CARLA maps.
\item A comprehensive and extensive evaluation of state-of-the-art VI systems using IBISCape sequences collected in dynamically simulated large-scale environments.
\end{itemize}

\setcounter{figure}{0}
\begin{figure}[tp]
\centering
\includegraphics[width=\linewidth]{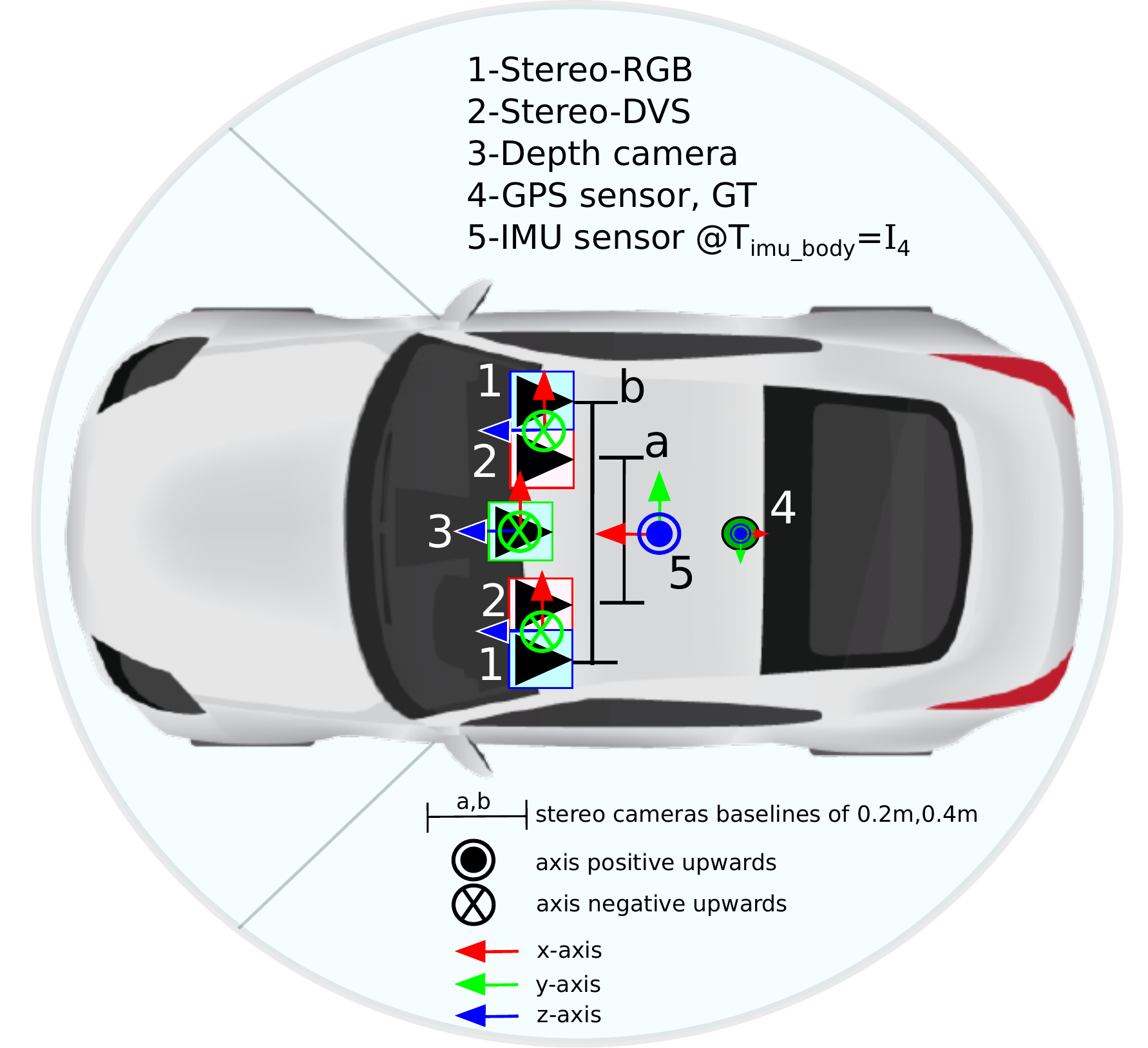}
\caption{Full sensor setup CAD model (Top view). The GPS readings are axis-aligned with the ground truth (GT). The IMU sensor frame is the vehicle body frame of reference with an identity transformation between them $I_{4\times4}$.\label{fig:IBISCape}}
\end{figure}

\setcounter{figure}{1}
\begin{figure*}[htbp]
\centering
\includegraphics[width=0.75\textwidth]{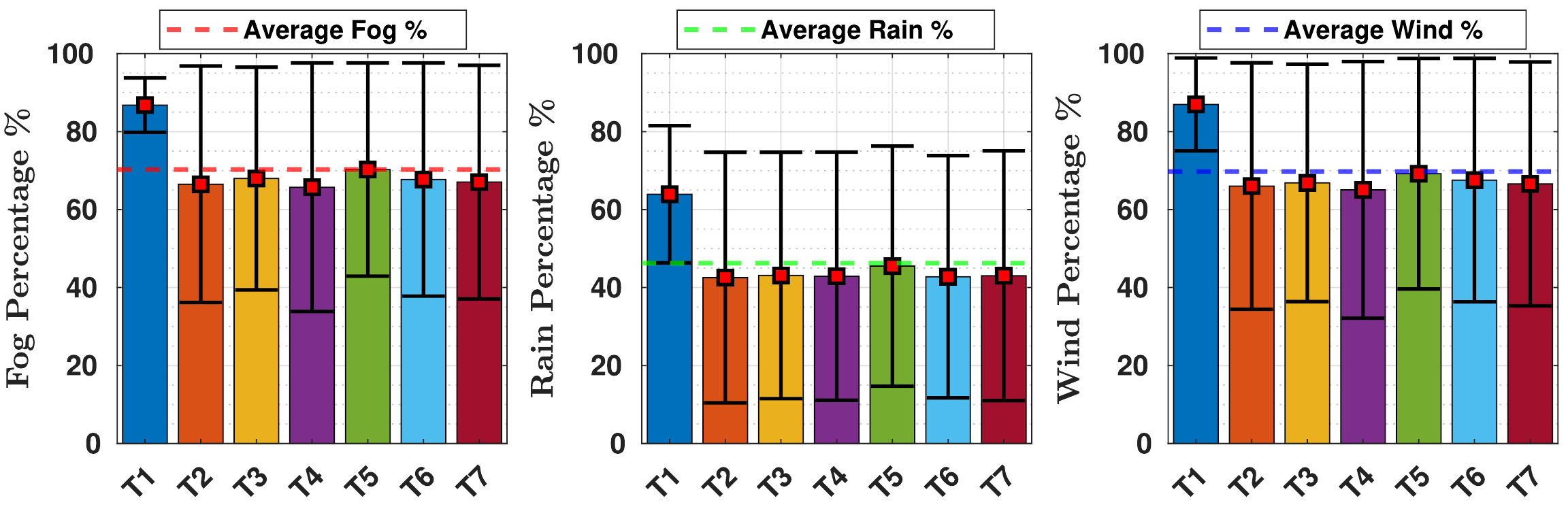}
\caption{The average fog, rain, and wind percentages for sample IBISCape sequences simulated in dynamic weather. $Ti$ is CARLA map of Town number (i). The percentage ranges can be set in IBISCape APIs within the weather simulation class.\label{fig:weather}}
\end{figure*}

This article is organized as follows: in Section \ref{sec:related}, we discuss the advantages and novelty of our benchmark compared to the related datasets in the field of multi-modal visual localization, including the state-of-the-art VI-SLAM algorithms. Section \ref{sec:core_sensor} explains the data acquisition APIs methodologies and the system calibration in details. Then, an extensive evaluation of the most recent VI systems using 25 IBISCape SLAM sequences with multiple modalities is represented in Section \ref{sec:eval}. Finally, in Section \ref{sec:conclusion}, we provide concluding remarks about our work including evaluation observations that motivate and push the development process of new multi-modal VI-SLAM techniques forward, especially in dynamic and large-scale environments based on new findings.

\section{Related Works\label{sec:related}}

\subsection{Existing Datasets}\label{sec:viodatasets}

The main goal of our benchmark’s data acquisition APIs is to collect multi-modal sequences suitable for most robotics perception evaluation, including scene understanding, calibration, and complete SLAM systems. IBISCape APIs are highly configurable concerning the intrinsic and extrinsic setup of the sensors and include all CARLA sensors till the version (0.9.11). 

Table \ref{table:comp_bench} compares the recent SLAM systems evaluation benchmarks from the sensors types and configuration point of view along with the carrier and ground truth information. Compared to the most recent publicly available benchmarks, IBISCape includes all the sensors needed to evaluate all the state-of-the-art VIO algorithms in any desired configuration including data rates and mono/stereo setups. 

Since IBISCape is a simulated benchmark, the GT data for the poses, vehicle controls, scene segmentation, and depth maps are rendered in high precision. This high precision GT data can significantly improve fitting the models of novel data-driven VIO architectures that lacks this high quality training data with the real-world datasets and hence improving the prediction accuracy.

\begin{sidewaystable*}
\sidewaystablefn%
\begin{center}
\begin{minipage}{\textheight}
\caption{Core Sensor Suite Comparison of Latest VIO Evaluation Benchmarks.}\label{table:comp_bench}%
\begin{tabular*}{\textheight}{@{\extracolsep{\fill}}lccccccc@{\extracolsep{\fill}}}
\toprule
\multicolumn{1}{l}{Benchmark Name} & {RGB} & {Depth} & {DVS\footnotemark[1]} & {Segmentation\footnotemark[2]} & {IMU} & {GT} & {Carrier} \\
\midrule
\multicolumn{1}{l}{TUM-RGBD \cite{sturm2012benchmark}}  & {Mono@30Hz} & {Mono@30Hz} &  {-} & {-} & {Accel@500Hz} & {MoCap@300Hz} & {Handheld\footnotemark[3]}\\
\multicolumn{1}{l}{KITTI \cite{geiger2013vision}}   & {Stereo@15Hz} & {-} &  {-} & {Annotations\footnotemark[4]} & {1@100Hz} & {GPS} & {Car}\\
\multicolumn{1}{l}{Malaga Urban \cite{blanco2014malaga}}   & {Stereo@20Hz} & {-} &  {-} & {-} & {1@100Hz} & {GPS} & {Car}\\
\multicolumn{1}{l}{UMich NCLT \cite{carlevaris2016university}}         & {Omni@5Hz} & {-} &  {-} & {-} & {1@100Hz} & {GPS/IMU/LiDAR} & {Segway}\\
\multicolumn{1}{l}{EuRoC \cite{burri2016euroc}}              & {Stereo@20Hz} & {-} &  {-} & {-} & {1@200Hz} & {Laser/Vicon} & {MAV}\\
\multicolumn{1}{l}{Zurich \cite{majdik2017zurich}}             & {Mono@30Hz} & {-} &  {-} & {-} & {1@100Hz} & {GPS} & {MAV}\\
\multicolumn{1}{l}{PennCOSYVIO \cite{pfrommer2017penncosyvio}}        & {Stereo@20Hz} & {-} &  {-} &    {-}   & {3@200Hz} & {Markers} & {Handheld}\\
\multicolumn{1}{l}{TUM-VI \cite{schubert2018vidataset}}                & {Stereo@20Hz} & {-} &  {-} &  {-} & {1@200Hz} & {MoCap} & {Handheld}\\
\multicolumn{1}{l}{Oxford \cite{judd2019oxford}}                & {Stereo@16Hz} & {Mono@30Hz} &  {-} &  {-}   & {1@500Hz} & {Vicon} & {Handheld}\\
\multicolumn{1}{l}{KAIST \cite{jeong2019complex}}                & {Stereo@10Hz} & {-} &  {-} &  {-}   & {1@200Hz} & {GPS} & {Car}\\
\multicolumn{1}{l}{OIVIO \cite{kasper2019benchmark}}                & {Stereo@30Hz} & {-} &  {-} &  {-}   & {1@100Hz} & {MoCap} & {Handheld}\\
\multicolumn{1}{l}{UZH-FPV \cite{delmerico2019we}}                & {Stereo@30Hz} & {-} &  {APS/DVS/IMU} &  {-}   & {1@500Hz} & {Laser} & {UAV}\\
\multicolumn{1}{l}{UMA-VI \cite{zuniga2020vi}}                & {Stereo@25Hz} & {-} &  {-} &  {-}  & {1@250Hz} & {Camera} & {Handheld}\\
\multicolumn{1}{l}{Blackbird \cite{antonini2020blackbird}}                & {Stereo@120Hz} & {Mono@60Hz} &  {-} &  {Mono@60Hz} & {1@100Hz} & {MoCap} & {UAV} \\
\multicolumn{1}{l}{VCU-RVI \cite{9341713}}                & {Mono@30Hz} & {Mono@30Hz} &  {-} &  {-} & {1@100Hz} & {MoCap} & {Handheld\footnotemark[3]} \\
\multicolumn{1}{l}{VIODE$^*$ \cite{9351597}}                & {Stereo@20Hz} & {-} &  {-} &  {Stereo@20Hz} & {1@200Hz} & {Simulation} & {UAV}\\
\multicolumn{1}{l}{EVENTSCAPE$^*$ \cite{gehrig2021combining}}                & {Mono@25Hz} & {Mono@25Hz} &  {Mono $\leq10^{6}e/s$} &  {Mono@25Hz}  & {-} & {Simulation} & {Car} \\
\multicolumn{1}{l}{TUM-VIE \cite{klenk2021tumvie}}                & {Stereo@20Hz} & {-} &  {Stereo $\leq10^{9}e/s$} &  {-}   & {1@200Hz} & {MoCap} & {Helmet}\\
\midrule
\multicolumn{1}{l}{\textbf{IBISCape}$^*$ (ours)} & {Stereo@20Hz} & {Mono@20Hz} & {Stereo $\leq10^{7}e/s$} & {Mono@20Hz} & {3@200Hz} & {Sim./GPS @200Hz} & {Car} \\
\botrule
\end{tabular*}
\footnotetext[1]{e/s is DVS events per second.}
\footnotetext[2]{Segmentation frames classify any visible object by displaying it in a different color according to its label (for example, pedestrians in a different color than cars). At the beginning of the simulation, each scene element is created with a tag. In the CARLA simulator, there are 23 segmentation tags with the possibility of adding new tags \url{https://carla.readthedocs.io/en/latest/tuto_D_create_semantic_tags/}.}
\footnotetext[3]{Some sequences where collected using a Robot for SLAM systems evaluation.}
\footnotetext[4]{Annotations for the dynamic objects in the scene are generated using scripts.}
\footnotetext[*]{Simulated benchmark.}
\end{minipage}
\end{center}
\end{sidewaystable*}

As an overview of the capabilities of the IBISCape benchmark, we collect simulated sequences on a car equipped with most of the low-cost sensors that can be used in the field of robotics perception. This simulation is thoroughly controlled by an autopilot that navigates the car on traffic-aligned roads. Furthermore, weather and scene constituents, including other cars and pedestrians, can be autonomously controlled within our APIs, resulting in datasets that can contend with the real-world benchmarks in the literature.

\subsection{Dynamic Environment Simulation}

 Minoda et al. \cite{9351597} introduce the concept of dynamic scene simulation with moving vehicles. In our benchmark, we extrapolate the concept of dynamic scene simulation to an entire dynamic environment simulation. This simulation includes moving vehicles and pedestrians, as well as a weather class. The weather ticking function updates the weather states every CARLA world tick with a specific speed factor and update frequency. The weather states that can be controlled are clouds, rain, wind, fog, humidity intensity, and sun angles.

A particular observation from sample IBISCape sequences in Fig. \ref{fig:weather} is that our weather update algorithm generates dynamic weather with high-intensity fog, rain, and wind with average percentages of 70\%, 45\%, and 70\%, respectively. These dynamic weather conditions result in high trajectory estimation errors due to map loss using existing VIO algorithms. This observation is further verified in Section \ref{sec:eval} where we compare the trajectory estimation accuracy in diverse weather conditions.

These weather challenges motivate the development of new VIO techniques based on the hybridization of heterogeneous multi-modal sensors to complete the shortages in the map lost during navigation. In Tab. \ref{table:resolution}, a brief comparison regarding the scene dynamic class and the amount of information being processed is represented in the camera’s frame resolution for all benchmarks represented in Tab. \ref{table:comp_bench}. The dynamic level indicators ([C]lear/[M]oderate/[D]ynamic) in Tab. \ref{table:resolution}, represent the severity of the [W]eather constituents such as: rain, fog, wind and lack of luminosity besides indicating the amount and speed of moving objects in the [S]cene such as other vehicles and walking pedestrians.

\begin{table}[tp]
\begin{center}
\begin{minipage}{\linewidth}
\caption{Benchmarks Dynamic Scene Information.}\label{table:resolution}
\begin{tabular}{@{}lll@{}}
\toprule%
\multicolumn{1}{l}{Benchmark} & {CAM Resolution [px]} & {Level\footnotemark[1]$^{,}$\footnotemark[2]} \\
\midrule
\multicolumn{1}{l}{TUM-RGBD}  & {1$\times{640}\times{480}$} &{C}\\
\multicolumn{1}{l}{KITTI}   & {2$\times{1384}\times{1032}$} & {C} \\
\multicolumn{1}{l}{Malaga Urban}   & {2$\times{1024}\times{768}$} & {M} \\
\multicolumn{1}{l}{UMich NCLT}         & {1$\times{1600}\times{1200}$} & {D (W/S)} \\
\multicolumn{1}{l}{EuRoC}              & {2$\times{752}\times{480}$} & {D (S)} \\
\multicolumn{1}{l}{Zurich}             & {1$\times{1024}\times{768}$} & {M} \\
\multicolumn{1}{l}{PennCOSYVIO}        & {2$\times{752}\times{480}$} & {C} \\
\multicolumn{1}{l}{TUM-VI}                & {2$\times{1024}\times{1024}$} & {C} \\
\multicolumn{1}{l}{Oxford}             & {2$\times{1280}\times{960}$} & {M} \\
\multicolumn{1}{l}{KAIST}               & {2$\times{1600}\times{1200}$} & {M} \\
\multicolumn{1}{l}{OIVIO}             & {2$\times{1280}\times{720}$} & {C} \\
\multicolumn{1}{l}{UZH-FPV}             & {2$\times{640}\times{480}$} & {C} \\
\multicolumn{1}{l}{UMA-VI}             & {2$\times{1024}\times{768}$} & {C} \\
\multicolumn{1}{l}{Blackbird}             & {2$\times{1024}\times{768}$} & {C} \\
\multicolumn{1}{l}{VCU-RVI}             & {1$\times{640}\times{480}$} & {D (S)} \\
\multicolumn{1}{l}{VIODE}             & {2$\times{752}\times{480}$} & {D (S)} \\
\multicolumn{1}{l}{EVENTSCAPE}    & {1$\times{512}\times{256}$} & {C} \\
\multicolumn{1}{l}{TUM-VIE}    & {2$\times{1024}\times{1024}$} & {D (S)} \\
\multicolumn{1}{l}{\textbf{Ours}} & {2$\times{1024}\times{1024}$} & {D (W/S)} \\
\botrule
\end{tabular}
\footnotetext[1]{C: Clear, M: Moderate, D: Dynamic.}
\footnotetext[2]{W: Weather, S: Scene.}
\end{minipage}
\end{center}
\end{table}

\subsection{Visual Odometry Techniques}

The novel VI systems are divided into two prominent techniques: loosely and tightly coupled fusion methodologies \cite{articlecompare}. In loosely coupled fusion \cite{autonomousproposed}, the camera is used as a black-box pose estimator \cite{7782863}, and an Extended Kalman Filter or an optimizer is applied to fuse the visual pose estimate with the pre-integrated noisy pose from IMU \cite{6696917}. Whereas in tightly coupled fusion, the scene descriptors (feature points) from the camera are directly inserted to the filter or optimizer to be fused with the IMU readings of the accelerometer and gyroscope using a model that estimates the pose, visual scale, IMU biases, and also re-project the optimized features to build a precise map of the scene.

The tightly coupled VI systems can be approached using two architectures: filter-based like MSCKF \cite{mourikis2007multi} and ROVIO \cite{bloesch2015robust}, and optimization-based such as VINS-Mono \cite{qin2018vins}, OKVIS \cite{leutenegger2015keyframe}, and recently ORB-SLAM3 \cite{Campos_2021} and BASALT \cite{usenko19nfr}. In the work of  Delmerico et al. \cite{8460664}, they compare all these VIO algorithms (except the recent works: ORB-SLAM3 and BASALT) in moderately constrained environments with respect to the dynamic level of the scene. They conclude that ROVIO and VINS-Mono are the best performing techniques concerning system latency, robustness, and accuracy.

In our article we focus on evaluating the most recent VI systems: BASALT and ORB-SLAM3 that share the same mapping layer concept based on ORB descriptors. However, their tracking architectures, IMU pre-integration methodologies, and loop-closing constraints are different. In Section \ref{sec:VIO_eval}, a qualitative performance analysis of BASALT and ORB-SLAM3 on multiple IBISCape SLAM sequences is performed. 

Since the DAVIS camera is a visual sensor with the highest temporal resolution (up to 1MHz) and dynamic range, it can be deemed one of the best sensors to deal with high speed robotics scenarios \cite{Zhou21tro} where conventional cameras may fail. Event cameras work on an unconventional caption technology based on the asynchronous detection of image intensity changes through all pixels on the retina. Novel open-source event-based VO algorithms have been developed in the last few years, including DVS monocular \cite{Gehrig_2020_CVPR} and stereo \cite{Zhou21tro} (ESVO) configurations. Using our APIs we generate stereo-DVS sequences with corner cases that are designed to push the limits of the event-based VO systems such as ESVO algorithm.


\section{Core Sensor Suite\label{sec:core_sensor}}

\setcounter{figure}{2}
\begin{figure*}[htbp]
\centering
\includegraphics[width=\textwidth]{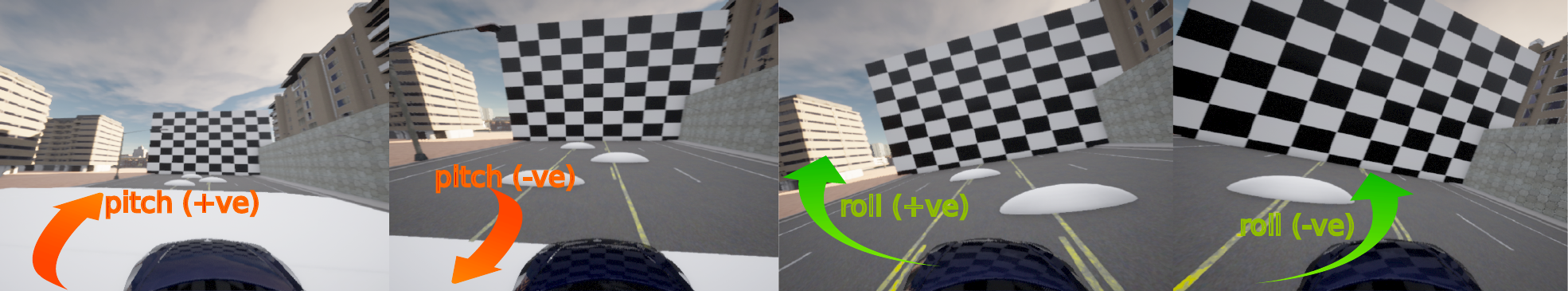}
\caption{Excitation of the vehicle pitch and roll angles using bubble bumps for reliable calibration results using Kalibr.\label{fig:excitationangles}}
\end{figure*}

\subsection{Data Acquisition Configuration \& Access}

Sensors in IBISCape APIs are highly configurable according to the intended mission, we have set an initial sensor configuration for our experiments that can be easily changed. This initial configuration of the IBISCape core sensor suite is given in Tab. \ref{table:comp_bench}. 

We generate data by eight acquisition APIs with four sensor setups mentioned in Tab. \ref{table:ibiscape_sequences} in two groups: 1. calibration and 2. SLAM. SLAM data acquisition APIs run on all CARLA maps with an autopilot for traffic-aligned navigation. On the other hand, calibration APIs run on our modified CARLA-map with manual vehicle control to apply desired motions to collect sequences with basic or complex motions. Both AprilGrid and Checkerboard targets are introduced during acquisition. Half of the calibration sequences are collected using the AprilGrid $6\times6$ and the other half using the Checkerboard $7\times7$.

In order to operate all sensors in the same acquisition API on multiple frequencies, we develop the following procedure: the core data acquisition concept is that the CARLA world clock ticks with the highest frequency sensor in the setup. After that, the system waits to listen to all sensors sending data at this tick, updates the weather conditions, and waits for a new world tick. This allows the acquisition of all sensors data with its occurrence timestamps. Then, one can apply any synchronization/calibration algorithms on the collected datasets as in \cite{yangicalib, leeefficient}. We apply this methodology (see Program \ref{prog:general}) to all sensor setups except the RGB-D setup, which requires time-synchronized and registered frames.

\begin{lstlisting}[language=Python, caption={Normal Data Acquisition API.},label={prog:general}]
data = []
sensor_list = Create_Sensors_List()
Create_Senors_Listener_handler(sensor_list)
Create_Weather_Control_class()
while (CARLA_world_tick()):
    Update_world_weather()
    for sensor in sensor_list:
        sensor.listen()
        if (RECORD_ON()):
            data.append(sensors.data())
\end{lstlisting}

On the contrary, the CARLA world ticks with the lowest frequency sensor in the RGB-D setup with CARLA \texttt{synchronous\_mode} acquisition (see Program \ref{prog:sync}). All the spawned sensors in the setup are stacked in a queue waiting for the world's tick to start listening to the data. Although all sensors operate with their frequencies, the API reads the measurements of all sensors simultaneously at the timestamp of that CARLA world tick.

\begin{lstlisting}[language=Python, caption={RGB-D Data Acquisition API.},label={prog:sync}]
data = []
sensors = Create_Sensors_List()
Create_Senors_Synchronization_Queue_class()
Create_Weather_Control_class()
while (CARLA_world_tick()):
    Update_world_weather()
    Sensors_Queue.tick(sensors)
    Sensors_Queue.listen(sensors)
    if (RECORD_ON()):
        data.append(sensors.data())
\end{lstlisting}

All the data acquisition APIs are executed on a 16 GB RAM laptop computer running 64-bit Ubuntu 20.04.3 LTS with AMD(R) Ryzen 7 4800h $\times{16}$ cores 2.9 GHz processor and a Radeon RTX NV166 Renoir graphics card.

Table \ref{table:ibiscape_sequences} shows the distribution of IBISCape sequences with different sensor modalities and configurations in all dynamic environmental conditions. All datasets in every sensor suite are synchronized during acquisition and timestamped in nano-seconds for high precision. Moreover, during the sequence collection, the vehicle control forces are saved as normalized vectors within the range [0, 1] and the steering angle ranges [-1, 1]. 

These control commands are normalized with respect to their maximum attained value based on the chosen vehicle dynamics. One of the advantages of CARLA simulator is that we can tune the physical properties of the vehicle and its wheels. 

Simulated GPS data is collected with all setups and synchronized with the GT pose. A text file with every framework explains its dataset files contents in detail. The data access manual for the 34 sequences and the acquisition APIs is given in details in the link in the supplementary information in Section \ref{secApndx1}.

\begin{table}[tp]
\begin{center}
\begin{minipage}{\linewidth}
\caption{IBISCape Sequences \& Sensor Setup.}\label{table:ibiscape_sequences}
\begin{tabular}{@{}ll|rrr@{}}
\toprule%
\multicolumn{2}{l|}{Acquisition Sensor Suite}& {Clear}& {Mod.}& {Dyn.} \\ 
\midrule
{\multirow{5}{*}{\begin{sideways}Calibration\end{sideways}}} & {IMU} & {1} & {-} & {-}\\
& {2xRGB+IMU (SVI)}& {2} & {-} & {-}\\
& {2xDVS+2xIMU (ESVI)}& {2} & {-} & {-} \\
& {RGB-D}& {2} & {-} & {-} \\
& {Full Sensor Setup}& {2} & {-} &  {-} \\
\midrule
{\multirow{4}{*}{\begin{sideways}SLAM\end{sideways}}} & {2xRGB+IMU (SVI)} & {2} & {2} & {3}\\
& {2xDVS+2xIMU (ESVI)}& {2} & {2} & {2} \\
& {RGB-D} & {2} & {2} & {2} \\
& {Full Sensor Setup} & {2} & {2} &  {2} \\
\midrule
\multicolumn{2}{c}{\textbf{Total = 34}} & {17} & {8} &  {9} \\
\botrule
\end{tabular}
\footnotetext{Note: All calibration evaluation sequences are collected in Clear weather only to achieve the best frame quality for robust and precise calibration results.}
\end{minipage}
\end{center}
\end{table}

\subsection{Cameras Intrinsic \& Extrinsic Calibration}

One of the advantages of IBISCape benchmark is providing calibration targets for evaluating multi-modal calibration algorithms as well as SLAM systems performance analysis. The more erroneous the calibration parameters, the more incorrect pose is estimated. Although the intrinsic calibration parameters of CARLA cameras can be configured directly in the APIs, there is no direct way to set the lens distortion coefficients till version (0.9.11). Consequently, we propose introducing the first calibration targets (Checkerboard ($7\times7$) and AprilGrid ($6\times6$)) to one CARLA map (Town 3). Moreover, to excite all angles, especially the pitch and roll angles which are not easy to be simulated in a car, we introduce artificial bumps in the form of bubbles and waves, as shown in Fig. \ref{fig:excitationangles}.

Furthermore, we use Kalibr \cite{rehder2016extending} to calibrate the stereo RGB cameras and the stereo DVS sensors after performing a frame reconstruction from events using the generic framework E2CALIB \cite{Muglikar2021CVPR}. Instead of simulating blinking leads that cannot be used in a multi-modal calibration framework (sample in Fig. \ref{fig:e2calib}). Since active illumination cannot be used to calibrate conventional cameras such as mono/stereo-RGB cameras, E2CALIB with the traditional calibration targets makes it possible to calibrate DVS sensors as any conventional camera. Hence, all cameras' intrinsic and extrinsic parameters in a multi-modal framework can be calibrated irrespective of their caption technology, i.e., frames or events.

\setcounter{figure}{3}
\begin{figure}[tp]
\centering
\includegraphics[width=\linewidth]{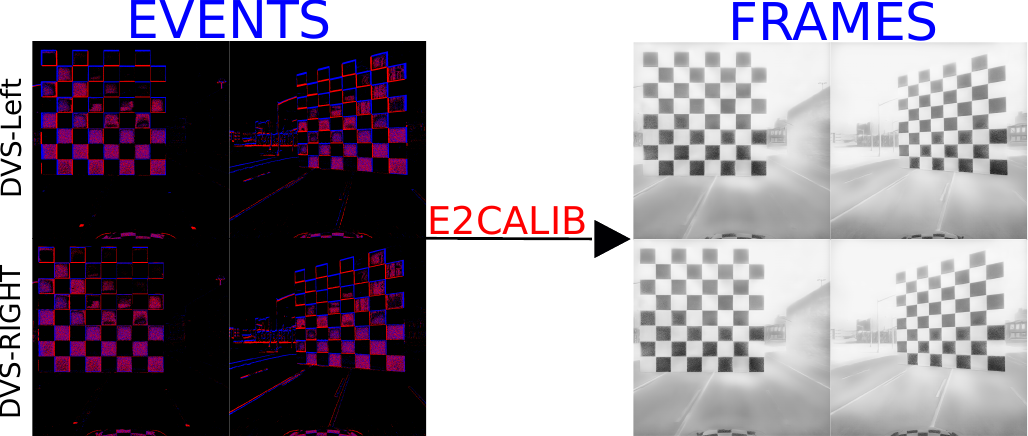}
\caption{Raw events to reconstructed frames using E2CALIB.\label{fig:e2calib}}
\end{figure}

All IBISCape cameras operate on a global shutter mode with a FOV of $90^{\circ}$. Table \ref{table:dvs} shows the specific DVS sensors parameters set during our simulations, including the positive/negative thresholds associated with an increment in brightness change along with their white noise standard deviation for positive/negative events.

The refractory period is the time lag between firing two consecutive events in nano-seconds. As a result, it limits the highest frequency of triggering events. The \texttt{log\_eps} value is used to convert RGB images to log-scale: L=log(\texttt{log\_eps}+\texttt{I\_gs}/255.0). Where (\texttt{I\_gs}) is the grayscale value of the RGB frame.

\begin{table}[tp]
\begin{center}
\begin{minipage}{\linewidth}
\caption{Simulated DVS Characteristics.}\label{table:dvs}
\begin{tabular}{@{}cc@{}}
\toprule%
\multicolumn{1}{c}{Parameter}  &  \multicolumn{1}{c}{Value set in CARLA}\\ 
\midrule
\multicolumn{1}{c}{\texttt{+ve/-ve\_threshold}} & \multicolumn{1}{c}{0.3} \\
\multicolumn{1}{c}{\texttt{sigma\_+ve/-ve\_threshold}} & \multicolumn{1}{c}{0.0} \\
\multicolumn{1}{c}{\texttt{refractory\_period\_ns}} & \multicolumn{1}{c}{0.0} \\
\multicolumn{1}{c}{\texttt{log\_eps}} & \multicolumn{1}{c}{0.05} \\
\botrule
\end{tabular}
\end{minipage}
\end{center}
\end{table}

The known camera model is a pinhole model with unknown distortion parameters for RGB and DVS cameras. We calibrate our cameras using Kalibr \textbf{pinhole-radial-tangential} and \textbf{pinhole-equidistant} distortion models. The calibration process is validated based on two criteria:
\begin{itemize}
    \item The estimated stereo baselines (extrinsics) compared to the GT values set in our acquisition APIs (see Tab. \ref{table:camerasintrinsics},\ref{table:baselines}).
    \item The quality of the optimization process that can be determined from the pixels re-projection errors and the number of optimization constraints (see Tab. \ref{table:reprojectionerrors}).
\end{itemize}

\begin{table*}[htbp]
\begin{center}
\begin{minipage}{\linewidth}
\caption{Stereo DVS sensors and RGB Cameras intrinsic parameters estimation using Kalibr. $f_x$ and $f_y$, $c_x$ and $c_y$ are the focal lengths and principal point coordinates, respectively. $k_1, k_2$ and $k_3, k_4$ are the radial and tangential distortion coefficients, respectively. Calibration is performed using the \textbf{Checkerboard} target.}\label{table:camerasintrinsics}
\begin{tabular}{@{}ll|rrrr|rrrr@{}}
\toprule%
\multicolumn{2}{l|}{Camera Model} & {$f_x$} & {$f_y$} & {$c_x$} & {$c_y\quad$} & {$k_1$} & {$k_2$} & {$k_3$} & {$k_4\quad$}\\
\midrule
{\multirow{4}{*}{\begin{sideways}DVS\end{sideways}}} & {cam0-radtan } & {517.07} & {517.59} &  {506.41} & {513.27\quad} & {-2.32e-3} & {7.12e-4} &  {1.97e-4} & {-8.87e-4\quad}\\
{}& {cam1-radtan }   & {517.45} & {517.79} &  {504.48} & {512.89\quad} & {-8.34e-4} & {-1.08e-3} &  {9.11e-5} & {-1.28e-3\quad}\\\cline{2-10}
\\[-1em]
{}& {cam0-equi} & {375.85} & {373.29} &  {573.79} & {513.44\quad} & {-0.0122} & {1.9684} &  {-3.8539} & {2.82\quad}\\
{}& {cam1-equi}    & {370.65} & {368.16} &  {572.65} & {513.1\quad} & {0.2912} & {0.2954} &  {-0.2626} & {0.2344\quad}\\
\midrule
\multicolumn{2}{c|}{\textbf{GT\quad cam0/cam1}} & {512.0} & {512.0} &  {512.0} & {512.0\quad} & {-} & {-} &  {-} & {-\quad}\\
\midrule
{\multirow{4}{*}{\begin{sideways}RGB\end{sideways}}}     &{cam0-radtan }       & {513.55} & {513.07} &  {511.0} & {510.26\quad} & {1.92e-3} & {-1.83e-3} &  {-8.5e-4} & {2.1e-4\quad}\\
{}& {cam1-radtan }   & {512.51} & {512.87} &  {512.0} & {512.1\quad} & {-2.75e-3} & {3.16e-3} &  {3.7e-4} & {-3.8e-4\quad}\\\cline{2-10}
\\[-1em]
{}& {cam0-equi}  & {511.11} & {511.18} &  {511.24} & {511.0\quad} & {0.3533} & {0.065} &  {0.181} & {-0.058\quad}\\
{}& {cam1-equi}    & {512.41} & {512.31} &  {512.0} & {512.34\quad} & {0.3269} & {0.1084} &  {0.1495} & {-0.0505\quad}\\
\midrule
\multicolumn{2}{c|}{\textbf{GT\quad cam0/cam1}} & {512.0} & {512.0} &  {512.0} & {512.0\quad} & {-} & {-} &  {-} & {-\quad} \\
\botrule
\end{tabular}
\end{minipage}
\end{center}
\end{table*}

\begin{table}[htbp]
\begin{center}
\begin{minipage}{\linewidth}
\caption{Estimation quality is further validated by comparison to the stereo baselines set in CARLA. cam0 and cam1 are the left and right cameras, respectively.}\label{table:baselines}
\begin{tabular}{@{}ll|r@{}}
\toprule%
\multicolumn{2}{l|}{Camera Model}& \multicolumn{1}{c}{Stereo Baseline (t [m])}\\ 
\midrule
{\multirow{4}{*}{\begin{sideways}DVS\end{sideways}}} & {cam0-radtan\quad} & {q=[3.18e-4  -1.77e-3  3.17e-5  1]} \\
{}& {cam1-radtan} & {t=[\textbf{-0.1986}  0.0009  0.0131]} \\\cline{2-3}
\\[-1em]
{}& {cam0-equi} & {q=[3.69e-4 -1.16e-3  -1.14e-4 1]}\\
{}& {cam1-equi} & {t=[\textbf{-0.1902}  0.003  0.0115]}\\
\midrule
\multicolumn{2}{c|}{\textbf{GT\quad cam0/cam1}} & {q=[0 0 0 1], t=[\textbf{-0.2} 0 0]}\\
\midrule
{\multirow{4}{*}{\begin{sideways}RGB\end{sideways}}} & {cam0-radtan}  & {q=[-0.0021 1.45e-4 4e-5 1]} \\
{}& {cam1-radtan}& {t=[\textbf{-0.403} 0.0103 -0.004]}\\\cline{2-3}
\\[-1em]
{}& {cam0-equi} & {q=[-.0014 -3.5e-4 -2e-5 1]}\\
{}& {cam1-equi} & {t=[\textbf{-0.413}  0.005  0.01]}\\
\midrule
\multicolumn{2}{c|}{\textbf{GT\quad cam0/cam1}} & {q=[0 0 0 1], t=[\textbf{-0.4} 0 0]}\\
\botrule
\end{tabular}
\end{minipage}
\end{center}
\end{table}

Based on these two criteria and the obtained results, we can conclude that the \textbf{pinhole-radtan} camera-distortion model best fits both RGB and DVS cameras simulation in CARLA. This conclusion is due to its lowest re-projection errors and highest stereo baseline estimation accuracy.

We provide all the calibration configuration files and various ROS scripts to convert the raw dataset files to \texttt{rosbag} and \texttt{.h5} file formats for Kalibr and E2CALIB frameworks.

\begin{table}[htbp]
\begin{center}
\begin{minipage}{\linewidth}
\caption{Re-projection errors \& optimization constraints.}\label{table:reprojectionerrors}
\begin{tabular}{@{}ll|rr@{}}
\toprule%
\multicolumn{2}{l|}{Camera Model}  & {Re-projection errors [px.]} & {Edges}\\ 
\midrule
{\multirow{4}{*}{\begin{sideways}DVS\end{sideways}}} & {cam0-radtan\quad} & {[0.000132, -0.000016]} & {61397} \\
{}& {cam1-radtan} & {[0.000163, -0.000009]} & {61397} \\\cline{2-4}
\\[-1em]
{}& {cam0-equi} & {[-0.000740, 0.000008]} & {61397} \\
{}& {cam1-equi} & {[-0.000703, 0.002294]} & {61397} \\
\midrule
{\multirow{4}{*}{\begin{sideways}RGB\end{sideways}}} & {cam0-radtan} & {[-0.000034, -0.000007]} & {29008} \\
{}& {cam1-radtan} & {[0.000034, 0.000007]} & {29008} \\\cline{2-4}
\\[-1em]
{}& {cam0-equi} & {[-0.000067, 0.000001]} & {29008} \\
{}& {cam1-equi} & {[0.000064, 0.000007]} & {29008} \\
\botrule
\end{tabular}
\end{minipage}
\end{center}
\end{table}

\subsection{Simulated IMU Calibration}

IBISCape novel calibration methodology is based on fixing the high quality calibration target in the center of the frame and moving the vehicle towards it with a complete manual control. Furthermore, adding bumps in its way in the form of a big wave and spherical bubbles, can ensure the sufficient excitation of the inertial sensor for precise system (IMU+cameras) calibration.

In CARLA, IMU measurements are modeled as most low-cost real-world IMUs containing a particular bias $b$ and White Gaussian noise $n$. Thus, the GT angular velocities $\omega$ and linear accelerations $a$ in the IMU frame are modeled as

\begin{equation}
\begin{array}{cc}
{\omega}_{GT}={\omega}_{gyro}-b_{g}-n_{g}, & a_{GT}=a_{accel}-b_{a}-n_{a}.\end{array}
\end{equation}

The standard deviation $\sigma_{wa},\sigma_{ba},\sigma_{wg},\sigma_{bg}$ values are given in Tab. \ref{table:imuvalues}, and Allan Deviation plots are given in Fig. \ref{fig:imucalibration} calibrated by the IMU Still Calibration Tool in \cite{zuniga2020vi} using a 168 [hrs] of IMU simulated sequence.

In Tab. \ref{table:imuvalues}, IMU still calibration shows a remarkable difference between the GT values we set in CARLA and the estimated ones. Besides, till CARLA version 0.9.11, the acceleration bias standard deviation value cannot be manually set within the simulation. As a result, an accurate and reliable IMU still calibration is essential to obtain simulated datasets with usable IMU measurements. We evaluate the IBISCape Stereo-Visual Inertial (SVI) sequences using the calibrated values for the IMU noises. 

\begin{table}[bp]
\begin{center}
\begin{minipage}{\linewidth}
\caption{Simulated IMU Still Calibration Results.}\label{table:imuvalues}
\begin{tabular}{@{}l|rr@{}}
\toprule%
\multicolumn{1}{l|}{Parameter}  & {CARLA} & {Calibration Tool}\\ 
\midrule
{$\sigma_{wa}$ [$m/s^2/\sqrt{Hz}$]} & {0.07} & {0.1103630988} \\
{$\sigma_{ba}$ [$m/s^3/\sqrt{Hz}$]} & {N/A} & {0.0126311958} \\
{$\sigma_{wg}$ [$rad/s/\sqrt{Hz}$]} & {0.004} & {0.0130318766} \\
{$\sigma_{bg}$ [$rad/s^2/\sqrt{Hz}$]\quad} & {0.0} & {0.0007030289} \\
\botrule
\end{tabular}
\end{minipage}
\end{center}
\end{table}

\setcounter{figure}{4}
\begin{figure}[bp]
\centering
\includegraphics[width=\linewidth]{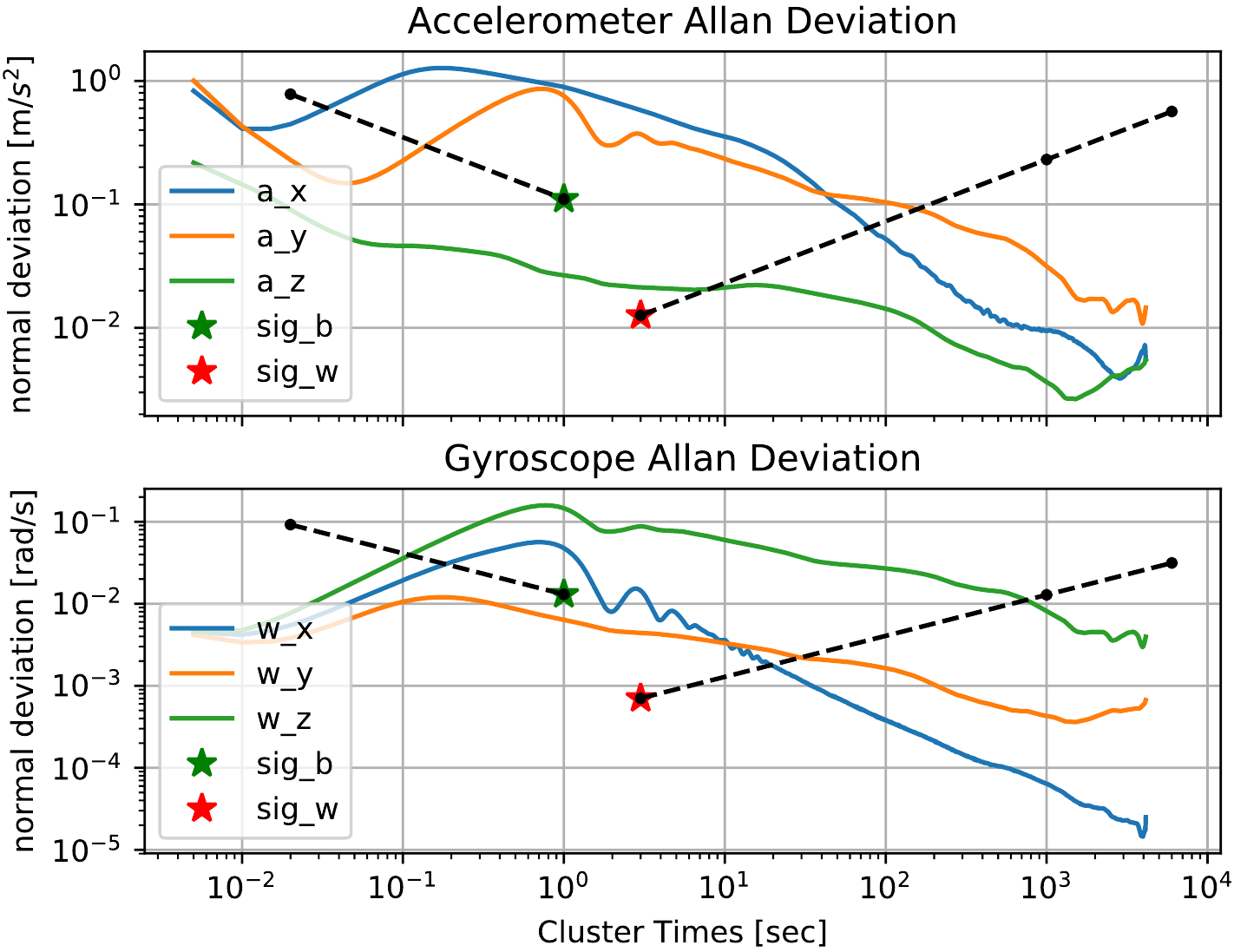}
\caption{IMU log-log scaled plot of Allan-variances over the cluster time. We calculate the IMU noises $\sigma_b$ and $\sigma_w$ at cluster times 1 sec. and 3 sec. with slopes $\mp$1/2 respectively.\label{fig:imucalibration}}
\end{figure}

\subsection{Inter-sensor Extrinsic Parameters}

The CAD model of the GT extrinsic relation between all the sensors in a full sensor setup is represented in Fig. \ref{fig:CAD_front}. The axes shown on the camera's center-line are given for the visual sensors only: RGB, DVS, Depth, Segmentation. All IMUs axes conventions are similar to that shown on the IMU0 center-line. Axes color and direction conventions coincide precisely with the Top view CAD model in Fig. \ref{fig:IBISCape}.

There is no orientation change between cameras i.e. $\delta{\theta}=[0,0,0]$ and all cameras have the relative rotation $q_{cam_1}^{cam_2}=[0,0,0,1]$. In Tab. \ref{table:extrinsics}, we give the exact GT values for each sensor location with respect to the IMU0 (body) axes.

Note that in the RGB-D sensor setup, the simulated RGB and Depth cameras have a concentric configuration where both the focal centers are coincided. Moreover, IBISCape data acquisition APIs are written to be highly configurable with respect to the inter-sensor extrinsic parameters with the ease of adding and removing sensors.

\begin{table}[bp]
\begin{center}
\begin{minipage}{\linewidth}
\caption{IBISCape Full Sensor Setup Extrinsics.}\label{table:extrinsics}
\begin{tabular}{@{}l|c@{}}
\toprule%
\multicolumn{1}{l|}{Sensor} & \multicolumn{1}{c}{X,Y,Z Translation to IMU0 [m]}\\ 
\midrule
{Left RGB} & {[0.0, 0.2, -2.8]} \\
{Right RGB} & {[0.0, -0.2, -2.8]} \\
{left DVS + IMU1} & {[0.0, 0.1, -2.8]} \\
{Right DVS + IMU2\quad} & {[0.0, -0.1, -2.8]} \\
{RGB-D cameras} & {[0.0, 0.0, -2.8]} \\
{GPS} & {[0.2, 0.0, -2.8]} \\
{GT Segmentation} & {[0.0, 0.0, -2.8]} \\
{GT Pose} & {[0.0, 0.0, 0.0]} \\
\botrule
\end{tabular}
\end{minipage}
\end{center}
\end{table}

\setcounter{figure}{5}
\begin{figure}[bp]
\centering
\includegraphics[width=\linewidth]{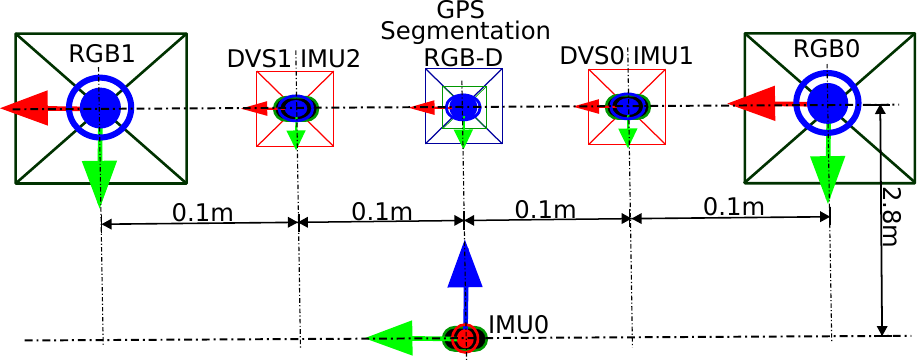}
\caption{Full sensor setup CAD model (Front view).\label{fig:CAD_front}}
\end{figure}

\section{Evaluation\label{sec:eval}}

\subsection{Efficient VI Systems\label{sec:VIO_eval}}
We use our IBISCape sequences to evaluate state-of-the-art monocular and stereo VI-SLAM algorithms which are ORB-SLAM3 and BASALT VIO. Their choice is because they are the latest state-of-the-art SLAM (ORB-SLAM3) and VIO (BASALT) algorithms. Accordingly, their extensive evaluation on new large-scale and dynamic environment (scene and weather) IBISCape sequences can facilitate detecting their limitations and performance regarding their accuracy and robustness.

BASALT uses a sparse set of FAST keypoints, tracks them between consecutive frames based on optical flow, and uses a pyramidal resolution method to ensure reliable and robust tracking in large-scale displacements tracking. Two layers for local bundle adjustment and global pose graph optimization are implemented for precise localization, mapping, and loop-closing. Furthermore, partial marginalization non-linear factors are applied to remove the IMU and feature outlier measurements for constant latency localization.

ORB-SLAM3 is developed to withstand a prolonged duration of low visual information. When a map is disturbed, it initiates a new map that will be smoothly merged with previous maps when revisiting similarly mapped areas. That results in a robust system that operates in dynamic environments and is much more accurate and robust than previous approaches.

Both ORB-SLAM3 and BASALT relate to the optimization-based tightly-coupled fusion stereo VI systems. In Section \ref{sec:SVI}, a detailed evaluation of their performance in large-scale dynamic environments.

The evaluation process is run on \texttt{25 IBISCape sequences} simulated in various large-scale dynamic environments. The GT length, duration, and the number of frames specifications of IBISCape sequences for all the data acquisition setups are given in Tab. \ref{table:sequences_desc}.

\begin{table}[htbp]
\begin{center}
\begin{minipage}{\linewidth}
\caption{IBISCape Sequences Specifications.}\label{table:sequences_desc}
\begin{tabular}{@{}lrrr@{}}
\toprule%
\multicolumn{1}{l}{\multirow{3}{*}{Sequence}} & \multicolumn{3}{c}{Specifications} \\
\cmidrule(lr){2-4}
& \multicolumn{1}{c}{Length} & \multicolumn{1}{c}{Duration} & {\multirow{2}{*}{Frames\footnotemark[1]}}\\
& \multicolumn{1}{c}{[m]} & \multicolumn{1}{c}{[sec]} & {} \\
\midrule
\multicolumn{1}{l}{Full Setup} & {}& {}& {} \\
\texttt{Clear-1} & {214.6313} & {60.52} & {1211}\\
\texttt{Clear-2} & {251.0401} & {70.55} & {1412}\\
\texttt{Moderate-1} & {368.9815} & {71.08}& {1422} \\
\texttt{Moderate-2} & {104.5391} & {29.92} & {599}\\
\texttt{Dynamic-1} & {217.9678} & {70.24} & {1405}\\
\texttt{Dynamic-2} & {61.2707} & {23.38} & {468}\\
\midrule
\multicolumn{1}{l}{SVI Setup} & {} & {} & {}\\
\texttt{Clear-1} & {140.2081} & {70.16} & {1404}\\
\texttt{Clear-2} & {141.1631} & {71.45} & {1429}\\
\texttt{Moderate-1} & {253.8933} & {64.40} & {1288}\\
\texttt{Moderate-2} & {330.6167} & {85.98} & {1719}\\
\texttt{Dynamic-1} & {248.6546} & {72.35} & {1448}\\
\texttt{Dynamic-2} & {289.0983} & {74.01} & {1480}\\
\texttt{Accident} & {23.6777} & {6.13} & {123}\\
\midrule
\multicolumn{1}{l}{RGB-D Setup} &{} & {} & {}\\
\texttt{Clear-1} & {223.1038} & {74.95} & {1500}\\
\texttt{Clear-2} & {360.5324} & {89.55} & {1792}\\
\texttt{Moderate-1} & {209.1469} & {72.65} & {1454}\\
\texttt{Moderate-2} & {233.6294} & {70.00} & {1401}\\
\texttt{Dynamic-1} & {208.0217} & {65.50} & {1311}\\
\texttt{Dynamic-2} & {406.3022} & {75.65} & {1514}\\ 
\midrule
\multicolumn{1}{l}{ESVI Setup} &  {} & {} & {}\\
\texttt{Clear-1} & {116.3213} & {23.98} & {4672}\\
\texttt{Clear-2} & {251.5679} & {60.61} & {12123}\\
\texttt{Moderate-1} & {264.8653} & {72.91}& {13980} \\
\texttt{Moderate-2} & {274.8627} & {61.13} & {4390}\\
\texttt{Dynamic-1} & {333.2455} & {71.54} & {13997}\\
\texttt{Dynamic-2} & {15.0866} & {23.87} & {11771}\\
\botrule
\end{tabular}
\footnotetext[1]{The number of RGB/Event frames in the sequence.}
\end{minipage}
\end{center}
\end{table}

\subsection{Performance Analysis}
We perform the ORB-SLAM3 and BASALT performance analysis using two evaluation metrics: 
\begin{enumerate}
\item[(i)] Absolute Trajectory Error (ATE) defined in \cite{sturm2012benchmark} for all (n) estimated poses as: 

\begin{equation}
    \textrm{ATE($\hat{T}^{(1:n)},T_{gt}^{(1:n)}$)} = \sqrt{\frac{1}{n}\sum_{i=1}^{n}{{\left\lvert\left\lvert t_i\right\rvert\right\rvert}^2}}\hspace{1mm}[m],
\end{equation}

where $\hat{T}^{(1:n)}$, $T_{gt}^{(1:n)}\in{SE(3)}$ are the estimated and ground truth trajectories, respectively. $t_i\in{R(3)}$ is the translation vector of the absolute trajectory error $E_i$ at time step $i$ where $E_i(R_i,t_i) = T_{gt(i)}^{-1}T_{rel}\hat{T_i}\in{SE(3)}$, and $T_{rel}$ is rigid-body transformation corresponding to the least-squares solution that maps the $\hat{T}$ trajectory onto the $T_{gt}$ trajectory calculated by optimization.

\item[(ii)] Relative Pose Error (RPE) at every i-th frame RPE is also defined in \cite{sturm2012benchmark} as: 

\begin{equation}
    \textrm{RPE($\hat{T}^{(1:n)},T_{gt}^{(1:n)}$)} = {\left\lvert\left\lvert \delta{t_i}\right\rvert\right\rvert\quad[m],}
\end{equation}

where $\delta{t_i}$ is the translation vector of the relative pose error $e_{i}(\delta\theta_i,\delta{t_i}) = (T_{gt(i)}^{-1}T_{gt(i+\Delta)})^{-1}(\hat{T}_{(i)}^{-1}\hat{T}_{(i+\Delta)})\in{se(3)}$ at time step $i$ with a fixed time interval $\Delta$ for our local trajectory increments.

For the orientations RPE values are given in degrees, we use the same formula after replacing the translation vector $\delta{t_i}$ with the rotation part $\delta\theta_i$ in $e_i$ by applying the $vee$ operator to the skew-symmetric error matrix:

\begin{equation}
    \textrm{RPE($\hat{T}^{(1:n)},T_{gt}^{(1:n)}$)} = {\left\lvert\left\lvert \left\lfloor \delta\theta_i\right\rfloor _{\vee}\right\rvert\right\rvert\quad[rad]}
\end{equation}

\end{enumerate}

We discuss a thorough descriptive and analytical evaluation for the ORB-SLAM3 and BASALT systems in the following sub-sections. The descriptive and analytical studies for every sensor setup raise the confidence in the novelty and usability of the IBISCape benchmark, using the calibrated RGB and DVS cameras distortion parameters along with the IMU still calibration.

\subsubsection{SVI Setup Evaluation}\label{sec:SVI}

IBISCape Stereo Visual Inertial (SVI) sequences push one of the limits of the ORB-SLAM3 system as mentioned in \cite{Campos_2021}, which is the IMU initialization of planar motion of vehicles like cars. In Fig. \ref{fig:svi_rgbd_setup}(A), this limitation constraint was further tested using the \texttt{Dynamic 1} sequence with significantly dimmed light and rapid scene motions. The ORB-SLAM3 IMU initialization failed to start with the mapping layer. This failure has led to a significant trajectory drift due to the map loss. This IMU initialization failure problem is also observed in the \texttt{Dynamic 2} sequence with the BASALT system.

In Tab. \ref{table:evaluation}, the other sequences, \texttt{Clear 1,2, Moderate 1,2}, show superior performance for the trajectory estimation using the ORB-SLAM3 system over BASALT based on both overall ATE and incremental RPE values. IBISCape SVI sequences are provided in \textbf{raw} and \textbf{rosbag} formats, along with the evaluation configuration files \texttt{.json} and \texttt{.yaml} for BASALT and ORB-SLAM3.

Although sharing ORB keypoints for loop-closing in BASALT and scene descriptors in ORB-SLAM3, BASALT has shown superior accuracy and robustness regarding the visual-inertial sub-system than an early version of ORB-SLAM \cite{usenko19nfr}. This better performance is due to the inertial layer of BASALT that utilizes recovered non-linear factors summarizing IMU and visual tracking on the higher layer of VIO.

However, the latest version of ORB-SLAM3 proved to be much more accurate than BASALT during evaluation on most of the IBISCape sequences, as shown in the performance analysis results in the following sub-section. Despite the superior performance of ORB-SLAM3 over BASALT, we note that the trajectory estimation is much faster in BASALT than in ORB-SLAM3. This evaluation observation validates the proposed comparison in Tab. (I) in \cite{Campos_2021}.

\subsubsection{RGB-D Setup Evaluation}

One of the advantages of IBISCape sequences is the variety of its sensors' multi-modality. While SVI sequences can provide the scene depth information by stereo RGB cameras and augment the scale factor using the inertial measurements, IBISCape RGB-D sequences offer another sensor modality to measure the scene depth: the depth camera. After alignment with the GT and scale factor recovery using the GPS measurements, we evaluate two ORB-SLAM3 algorithms: the monocular RGB and the RGB-D SLAM systems. In Fig. \ref{fig:svi_rgbd_setup}, it is evident that adding the depth information results in more accurate trajectory estimation with a minor map loss in dynamic weather.

We notice this map loss clearly with the mono-RGB using \texttt{Dynamic 1,2, Moderate 1,2} sequences. However, in clear weather sequences \texttt{Clear 1,2}, the monocular RGB SLAM can outperform the RGB-D SLAM as seen in Tab. \ref{table:evaluation} with respect to the ATE values. IBISCape RGB-D sequences are provided in \textbf{raw} format with the RGB and Depth cameras \texttt{association.txt} file for every sequence, along with the evaluation configuration \texttt{.yaml} files for ORB-SLAM3 RGB-D and mono-RGB systems.

\subsubsection{ESVI Setup Evaluation}

IBISCape event-based sequences are designed to address two corner case scenarios introduced to the event-based monocular/stereo VO algorithms. The first scenario is the planar motion in \textbf{large-scale} environments; this scenario leads to the generation of millions of events fired at locations in the scene that can be tens of meters away from each other. These environments consume much time to reconstruct a map, leading to significant processing time gaps between the tracking and mapping layers of the odometry algorithm. As a result, all the ESVO \cite{Zhou21tro} experiments on IBISCape sequences failed during the trajectory estimation giving an error indicating inconsistency between the tracking and mapping layers, although maps initialize successfully.

The second scenario is the dynamic weather, including fog and rain droplets that can cause random asynchronous events. Hot pixels in real-world DVS can be hardware defects, but simulated DVS can indicate random rain/fog firings in CARLA. Applying a hot pixel filter can detect and remove these unexpected events. Figure \ref{fig:histo_hot_pixel} shows a sample of the hot pixels removed due to fog and rain. Removing hot pixels in DVS sensor is based on two criteria: the highest $N$ pixels firing most events, or the pixels firing greater than $n_{\sigma}\times\sigma$ events. Where $n_{\sigma}$ is event occurrence standard deviation multiplier and $\sigma$ is the event occurrence standard deviation.

\setcounter{figure}{7}
\begin{figure}[tp]
\centering
\includegraphics[width=\linewidth]{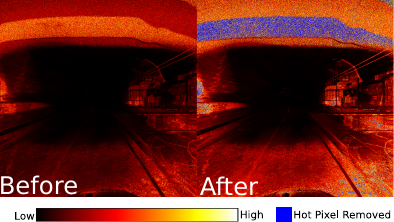}
\caption{Histogram of Events before and after the hot pixels removal with 26.39\% of events discarded, caused mainly by fog and rain puddles. Sample from \texttt{FULL\_Dynamic\_1} sequence.\label{fig:histo_hot_pixel}}
\end{figure}

In order to evaluate IBISCape stereo-DVS calibration parameters, we construct stereo-RGB frames from the events using the E2VID pre-trained model \cite{Rebecq19cvpr}. We assess the ORB-SLAM3 stereo RGB SLAM system on these reconstructed frames. Despite filtering the scene from noisy events resulting from fog and rain droplets, Table \ref{table:esvi_evaluation} shows a complete failure in trajectory estimation in the case of \texttt{Dynamic 1} sequence. Due to the dynamic weather conditions and rapid system dynamics E2VID frame reconstruction fails with most of IBISCape sequences. Consequently, accumulating event arrays to reconstruct high quality frames was the best alternative to evaluate ORB-SLAM3 stereo-RGB system on IBISCape ESVI sequences. Figure \ref{fig:comparison} represents a comparison on different frame reconstruction methods regarding features detection. 

The only case where the E2VID can evaluate event-based SLAM systems performance in trajectory estimations is clear weather with a low dynamic scene level as in the \texttt{Clear 1} sequence. A rapid change in the scene can cause an instantaneous map loss that affects the whole trajectory estimation even if the weather is clear, as in the \texttt{Clear 2} sequence. IBISCape ESVI sequences are provided in raw \texttt{.npz} (NumPy arrays) and bag formats for the stereo-DVS events, along with \texttt{timestamps.csv} file includes the start and end timestamps for every time surface. Besides, the E2VID grayscale frames reconstruction results and the locations of the hot pixels for the stereo-DVS are also provided.

\subsubsection{FULL Setup Evaluation}

The most significant contribution of the IBISCape benchmark is its FULL sensor setup sequences, where all sequences contain a combination of all the available sensors simulated in clear/moderate/dynamic weather environments. As a result, a complete comprehensive quantitative evaluation of all the SLAM systems mentioned in the previous sub-sections can be compared on the same sequence for every specific weather condition, as seen in Fig. \ref{fig:full_setup}. We represent in Tables \ref{table:evaluation} and \ref{table:esvi_evaluation} an extensive qualitative assessment of the state-of-the-art SLAM systems based on the six FULL setup sequences. Regarding \texttt{Clear 1,2}, the trajectory estimation is aligned with the groundtruth profile until a rapid motion occurs and the events map is disturbed. Each IBISCape FULL setup sequence is equipped with all the data formats as given with the specialized setup sequences. 

Based on all the evaluation observations, we can conclude that the current pre-trained models to reconstruct frames can be unreliable specially in dynamic weather and large-scale environments as represented in Fig. \ref{fig:comparison}. This gives the most important advantage of IBISCape benchmark providing thousands of event arrays collected in a way to ease the retraining of the current models and motivates the development of new approaches to process events in such scenarios and corner cases.

The most prominent conclusion from evaluations on the FULL setup is that in dynamic weather DVS sensor cannot be reliable to estimate the pose of the AGV. This conclusion is since events are fired asynchronously with high frequency, causing the visual sensor to be susceptible to weather constituents like rain or fog, which can degrade the estimation performance. Accordingly, our multi-modal datasets with the simulated corner cases can be the building block of choose-case scenarios for selecting the most efficient combination of multi-modal VI sensors for AGVs navigating in the most challenging environments.

\begin{table*}[htbp]
    \small\sf\centering
    \caption{ORB-SLAM3 (SVI, RGB-D, mono-RGB), BASALT, EVO and EMVS performance analysis based on both ATE and RPE evaluation metrics using IBISCape sequences in all simulated dynamic environments. Relative Pose Error (RPE) is formulated in terms of the mean $\pm$ standard deviation.\label{table:evaluation}}
    \begin{tabular*}{\textwidth}{@{\extracolsep{\fill}\quad}lcccccc}
      \toprule
      \multicolumn{1}{l}{\multirow{2}{*}{Sequence}} & \multicolumn{3}{c}{Algorithm 1} &
      \multicolumn{3}{c}{Algorithm 2} \\
      \cmidrule(lr){2-4} \cmidrule(lr){5-7}
      & \multicolumn{1}{c}{ATE [m]} & \multicolumn{1}{c}{RPE [m]} & \multicolumn{1}{c}{RPE [deg]} & \multicolumn{1}{c}{ATE [m]} & \multicolumn{1}{c}{RPE [m]} & \multicolumn{1}{c}{RPE [deg]}\\
      \midrule
      \multicolumn{1}{l}{FULL Setup - I} & \multicolumn{3}{c}{ORB-SLAM3 - SVI} & \multicolumn{3}{c}{BASALT}\\
      \texttt{Clear-1} & {\textbf{13.7184}} & {\textbf{0.2852$\pm$0.1765}} & {1.1677$\pm$1.2825} & {18.7082} & {0.3231$\pm$0.2047} & {\textbf{0.2675$\pm$0.5794}} \\
      \texttt{Clear-2} & {12.3043} & {0.1234$\pm$0.1437} & {0.7866$\pm$0.8635} & {\textbf{12.0170}} & {\textbf{0.0655$\pm$0.0746}} & {\textbf{0.1699$\pm$0.3937}} \\
      \texttt{Moderate-1} & {\textbf{32.8159}} & {\textbf{0.4248$\pm$0.0748}} & {\textbf{0.3838$\pm$0.3351}} & {49.9634} & {0.6076$\pm$0.2327} & {0.1420$\pm$0.7516} \\
      \texttt{Moderate-2} & {\textbf{3.7829}} & {\textbf{0.1596$\pm$0.1356}} & {0.6265$\pm$0.9143} & {11.8746} & {0.2190$\pm$0.1318} & {\textbf{0.1645$\pm$0.4232}} \\
      \texttt{Dynamic-1} & {17.2807} & {0.2584$\pm$0.1782} & {0.3845$\pm$0.6576} & {\textbf{16.6205}} & {\textbf{0.2433$\pm$0.1908}} & {\textbf{0.1222$\pm$0.3261}} \\
      \texttt{Dynamic-2} & {\textbf{4.9187}} & {\textbf{0.1801$\pm$0.1520}} & {\textbf{0.0730$\pm$0.0607}} & {9.3084} & {0.2031$\pm$0.1757} & {0.1231$\pm$0.5529} \\
      \midrule
      \multicolumn{1}{l}{FULL Setup - II} & \multicolumn{3}{c}{ORB-SLAM3 - RGB-D} & \multicolumn{3}{c}{ORB-SLAM3 - mono-RGB}\\
      \texttt{Clear-1} & {20.2653} & {0.3418$\pm$0.2003} & {\textbf{1.1702$\pm$1.2942}} & {\textbf{3.5142}} & {\textbf{0.3103$\pm$0.1804}} & {1.1723$\pm$1.2939} \\
      \texttt{Clear-2} & {\textbf{14.8820}} & {\textbf{0.1659$\pm$0.1276}} & {\textbf{0.7914$\pm$0.8535}} & {18.4484} & {0.1759$\pm$0.1825} & {0.8240$\pm$0.8695} \\
      \texttt{Moderate-1} & {\textbf{40.1021}} & {\textbf{0.4612$\pm$0.1130}} & {\textbf{0.4307$\pm$0.3410}} & {67.1074} & {0.4995$\pm$0.0667} & {0.4212$\pm$0.3720} \\
      \texttt{Moderate-2} & {\textbf{3.5969}} & {\textbf{0.2595$\pm$0.1471}} & {\textbf{0.6520$\pm$0.9115}} & {15.0772} & {0.3040$\pm$0.2212} & {0.7999$\pm$0.9769} \\
      \texttt{Dynamic-1} & {\textbf{11.5730}} & {\textbf{0.2048$\pm$0.1478}} & {0.3504$\pm$0.5771} & {22.2793} & {0.3090$\pm$0.2640} & {\textbf{0.3154$\pm$0.6171}} \\
      \texttt{Dynamic-2} & {\textbf{15.5917}} & {\textbf{0.2824$\pm$0.2806}} & {0.1101$\pm$0.0955} & {17.2632} & {0.3210$\pm$0.4717} & {\textbf{0.0626$\pm$0.0679}} \\
      \midrule
      \multicolumn{1}{l}{SVI Setup} & \multicolumn{3}{c}{ORB-SLAM3 - SVI} & \multicolumn{3}{c}{BASALT} \\
      \texttt{Clear-1} & {\textbf{3.1262}} & {\textbf{0.1145$\pm$0.0728}} & {0.2699$\pm$0.4272} & {12.2769} & {0.1859$\pm$0.0234} & {\textbf{0.0839$\pm$0.4300}} \\
      \texttt{Clear-2} & {\textbf{1.6666}} & {0.1076$\pm$0.0577} & {0.3646$\pm$0.4795} & {4.0626} & {\textbf{0.0514$\pm$0.0845}} & {\textbf{0.0919$\pm$0.2049}} \\
      \texttt{Moderate-1} & {\textbf{11.5160}} & {\textbf{0.0496$\pm$0.0913}} & {\textbf{0.1278$\pm$0.3385}} & {70.1406} & {0.1290$\pm$0.0712} & {0.1087$\pm$0.5707} \\
      \texttt{Moderate-2} & {\textbf{8.8561}} & {\textbf{0.3207$\pm$0.1681}} & {0.5831$\pm$0.7833} & {27.4657} & {0.3587$\pm$0.1527} & {\textbf{0.1519$\pm$0.3228}} \\
      \texttt{Dynamic-1} & {50.7355} & {0.2580$\pm$0.1351} & {\textbf{1.1411$\pm$1.0240}} & {\textbf{12.4161}} & {\textbf{0.1853$\pm$0.1512}} & {0.2324$\pm$0.4938} \\
      \texttt{Dynamic-2} & {\textbf{9.5503}} & {\textbf{0.1188$\pm$0.1445}} & {0.1338$\pm$0.4248} & {41.4773} & {0.2414$\pm$0.1277} & {\textbf{0.0921$\pm$0.5366}} \\
      \texttt{Accident} & {16.2158} & {\textbf{0.2916$\pm$0.1594}} & {0.8453$\pm$1.9645} & {\textbf{2.6652}} & {0.4169$\pm$0.1158} & {\textbf{0.4808$\pm$0.9041}} \\
      \midrule
      \multicolumn{1}{l}{RGB-D Setup} & \multicolumn{3}{c}{ORB-SLAM3 - RGB-D} & \multicolumn{3}{c}{ORB-SLAM3 - mono-RGB}\\
      \texttt{Clear-1} & {20.9667} & {\textbf{0.2370$\pm$0.1846}} & {\textbf{0.3830$\pm$0.5798}} & {\textbf{4.9536}} & {0.2522$\pm$0.1970} & {0.3842$\pm$0.5761} \\
      \texttt{Clear-2} & {5.9339} & {\textbf{0.3647$\pm$0.2238}} & {0.3788$\pm$0.6832} & {\textbf{0.8387}} & {0.3706$\pm$0.2227} & {\textbf{0.3297$\pm$0.6424}} \\
      \texttt{Moderate-1} & {\textbf{2.8882}} & {\textbf{0.2872$\pm$0.2419}} & {0.0718$\pm$0.0740} & {14.6609} & {0.2873$\pm$0.2562} & {\textbf{0.0290$\pm$0.0276}} \\
      \texttt{Moderate-2} & {\textbf{13.5358}} & {0.2353$\pm$0.1597} & {0.2610$\pm$0.6043} & {29.3680} & {\textbf{0.2207$\pm$0.1559}} & {\textbf{0.2473$\pm$0.6575}} \\
      \texttt{Dynamic-1} & {\textbf{8.7264}} & {0.2732$\pm$0.2626} & {\textbf{0.5988$\pm$0.8542}} & {15.1911} & {\textbf{0.2628$\pm$0.2426}} & {0.6079$\pm$0.8009} \\
      \texttt{Dynamic-2} & {\textbf{12.0050}} & {\textbf{0.4743$\pm$0.1710}} & {0.5558$\pm$0.5380} & {121.4955} & {0.6201$\pm$0.3161} & {\textbf{0.5496$\pm$0.4548}} \\ 
      \bottomrule
    \end{tabular*}
\end{table*}

\begin{table*}[htbp]
    \small\sf\centering
    \caption{The ATE and RPE evaluation metrics for assessing the performance of ORB-SLAM3 Stereo Visual Odometry system using E2VID reconstructed frames from IBSCape Stereo-DVS events in both Full and ESVI sequences in all simulated dynamic environments. Relative Pose Error (RPE) is formulated in terms of the mean $\pm$ standard deviation.\label{table:esvi_evaluation}}
    \begin{tabular*}{\textwidth}{@{\extracolsep{\fill}\quad}lcccccc}
      \toprule
      \multicolumn{1}{l}{\multirow{2}{*}{Sensor Setup}} & \multicolumn{3}{c}{FULL Setup - III} &
      \multicolumn{3}{c}{ESVI Setup} \\
      \cmidrule(lr){2-4} \cmidrule(lr){5-7}
      & \multicolumn{1}{c}{ATE [m]} & \multicolumn{1}{c}{RPE [m]} & \multicolumn{1}{c}{RPE [deg]} & \multicolumn{1}{c}{ATE [m]} & \multicolumn{1}{c}{RPE [m]} & \multicolumn{1}{c}{RPE [deg]}\\
      \midrule
      \multicolumn{1}{l}{Sequence} & \multicolumn{3}{c}{E2VID + ORB-SLAM3 - Stereo-RGB} & \multicolumn{3}{c}{E2VID + ORB-SLAM3 - Stereo-RGB}\\
      \texttt{Clear-1} & {84.7657} & {0.7384$\pm$0.8979} & {0.6746$\pm$0.5510} & {9.4286} & {0.9747$\pm$1.8231} & {0.0324$\pm$0.1110} \\
      \texttt{Clear-2} & {156.8587} & {0.2047$\pm$0.0446} & {0.3806$\pm$0.4775} & {121.0946} & {0.7865$\pm$10.2468} & {0.2974$\pm$2.5672} \\
      \texttt{Moderate-1} & {157.9537} & {1.3439$\pm$0.4314} & {0.3565$\pm$0.1739} & {79.8216} & {25.2979$\pm$43.9429} & {4.8829$\pm$15.0766} \\
      \texttt{Moderate-2} & {29.1791} & {1.2812$\pm$1.0903} & {0.1766$\pm$0.1263} & {62.2875} & {0.6262$\pm$5.3079} & {0.6011$\pm$0.4367} \\
      \texttt{Dynamic-1} & {\color{red}235.7885} & {0.4666$\pm$2.3055} & {0.0274$\pm$0.0733} & {35.6279} & {0.4821$\pm$1.8246} & {0.6281$\pm$0.4795} \\
      \texttt{Dynamic-2} & {52.1609} & {5.4907$\pm$6.1910} & {0.1587$\pm$0.1005} & {106.0616} & {3.8763$\pm$18.3749} & {1.9862$\pm$3.5864} \\
      \bottomrule
    \end{tabular*}
\end{table*}

\setcounter{figure}{6}
\begin{figure*}[htbp]
\centering
\includegraphics[width=0.897\textwidth]{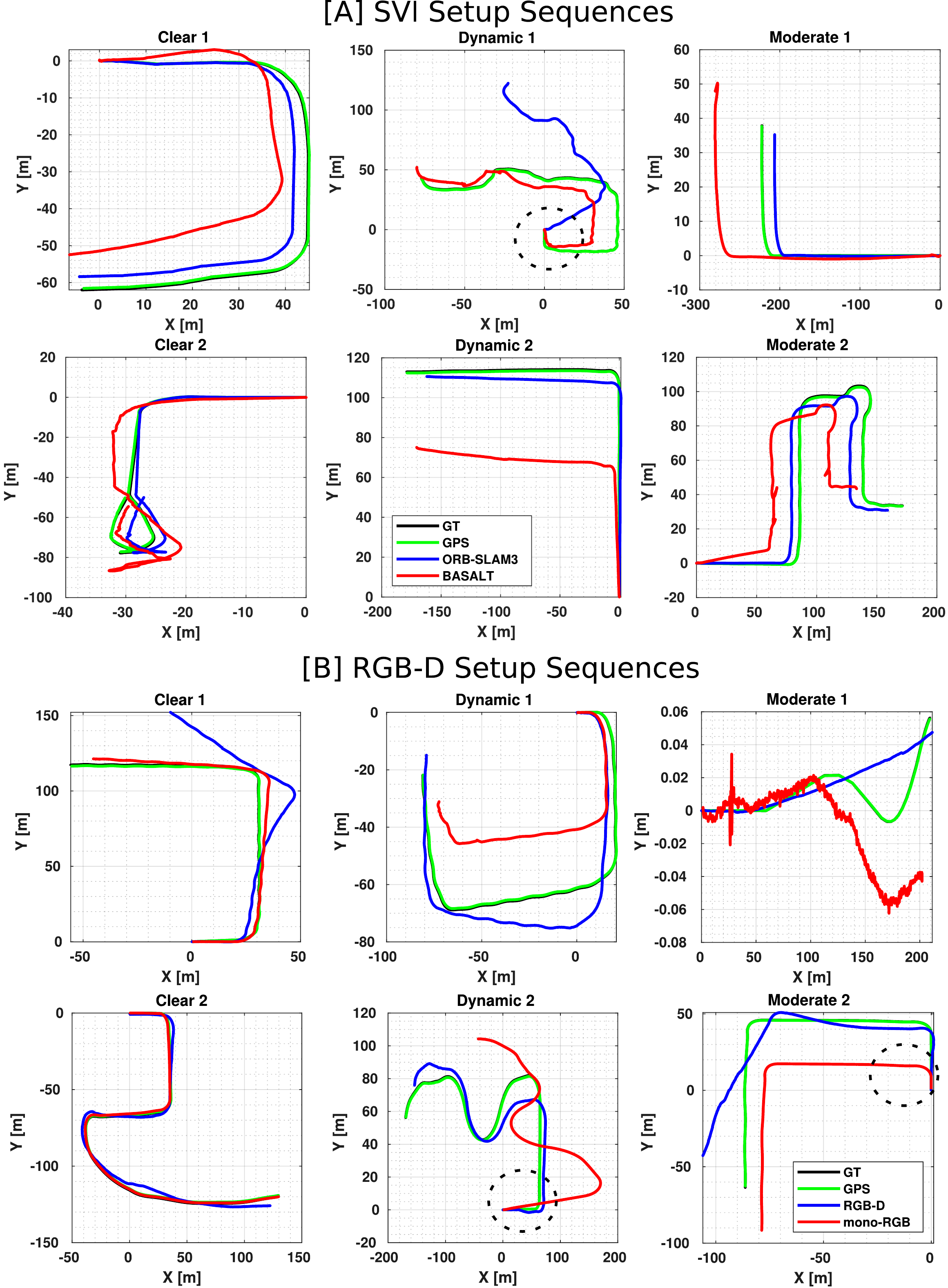}
\caption{Trajectories estimated by ORB-SLAM3 and BASALT SLAM systems using IBISCape sequences with \textbf{SVI sensor setup} and \textbf{RGB-D sensor setup}, with comparison to their ground truth and GPS paths. For the set \texttt{(A) SVI SETUP}, ORB-SLAM3 Stereo Visual Inertial Odometry (SVI) algorithm performance is analysed and compared to the BASALT SVI algorithm. Whereas for the set \texttt{(B) RGB-D SETUP}, two ORB-SLAM3 algorithms: Monocular RGB and RGB-D SLAM systems, are assessed with respect to each other after estimation alignment with the GT and scale factor recovery using the GPS measurements.\label{fig:svi_rgbd_setup}}
\end{figure*}

\setcounter{figure}{8}
\begin{figure*}[htbp]
\centering
\includegraphics[width=\textwidth]{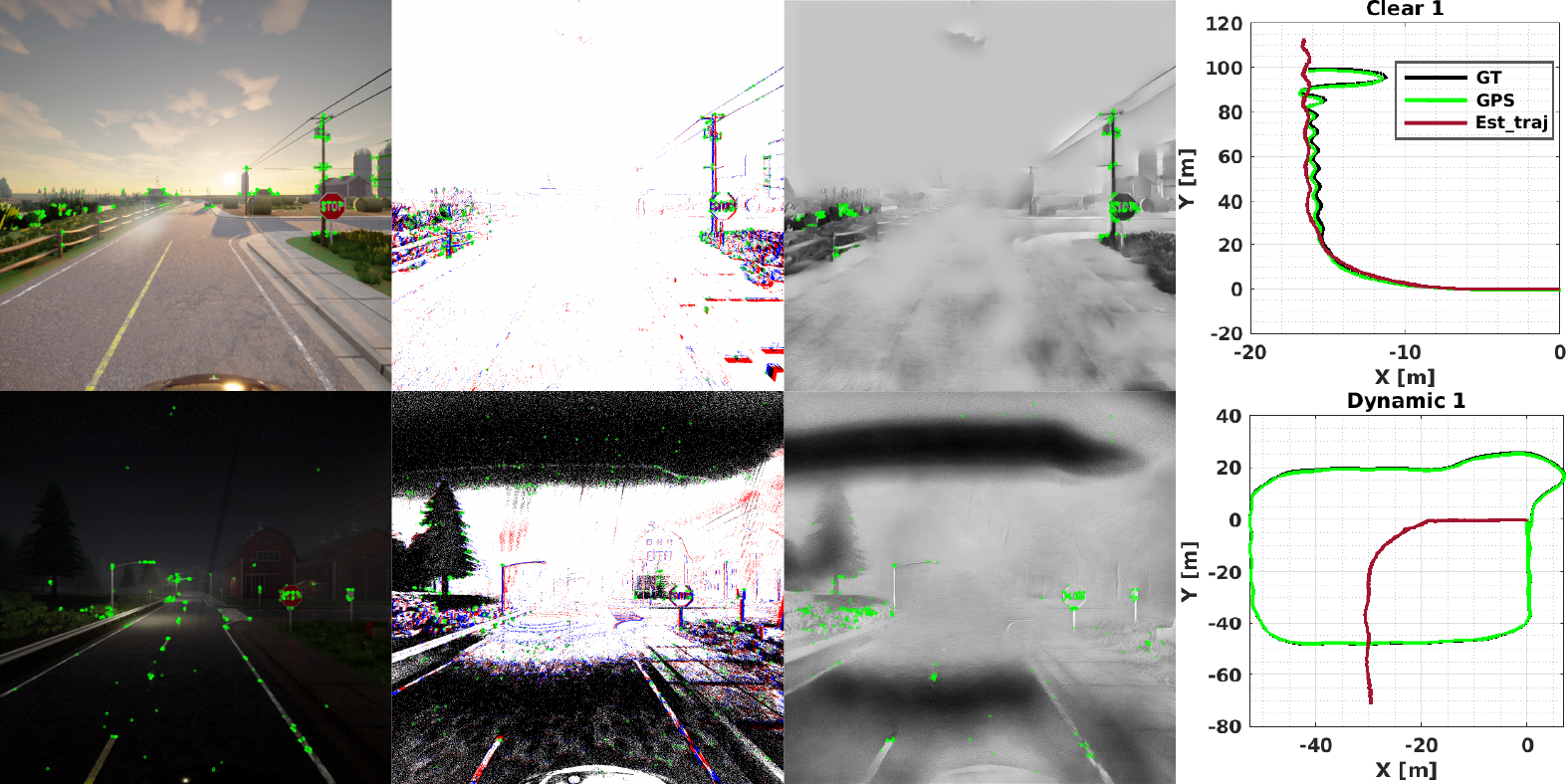}
\caption{The effect of dynamic weather conditions on DVS events and ORB feature extraction. From left to right: RGB image, accumulated events frame, E2VID reconstructed image and trajectory estimation based on accumulated events frames. All frames are simulated with the same frequency ($20Hz$). Both sequences \texttt{Clear 1, Dynamic 1} belong to the \textbf{ESVI sensor setup}. \label{fig:comparison}}
\end{figure*}

\setcounter{figure}{9}
\begin{figure*}[htbp]
\centering
\includegraphics[width=\textwidth]{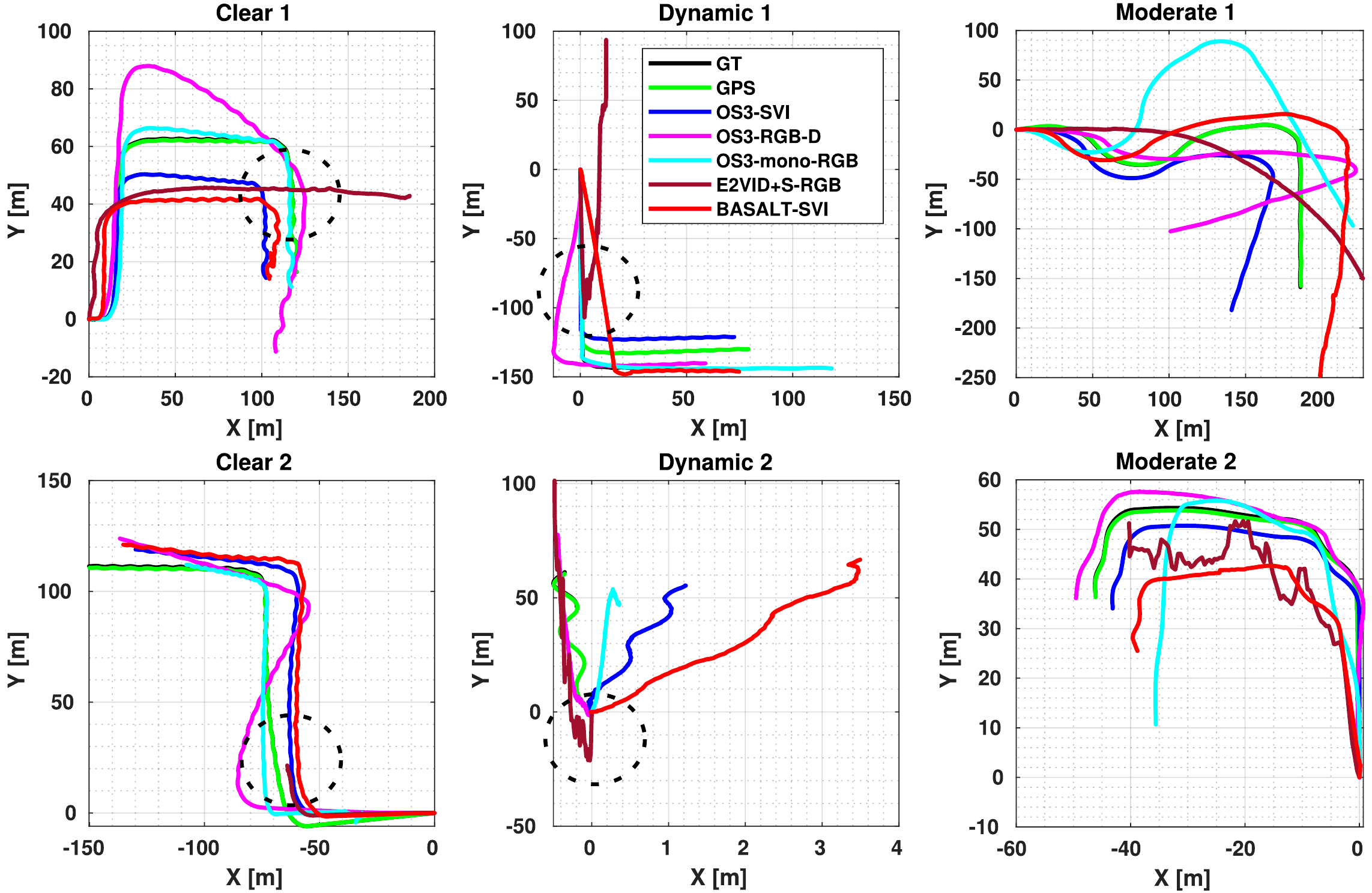}
\caption{Trajectories estimated by ORB-SLAM3 and BASALT SLAM systems using IBISCape sequences with \textbf{FULL sensor setup} and comparing to their ground truth and GPS paths. ORB-SLAM3 algorithms involved are: Monocular RGB, Stereo-RGB (S-RGB), Stereo Visual Inertial (SVI), and RGB-D SLAM systems. While for BASALT, the SVI algorithm is assessed.\label{fig:full_setup}}
\end{figure*}

\section{Conclusion\label{sec:conclusion}}
This article proposed the IBISCape simulated heterogeneous sensors benchmark in large-scale environments along with 34 sequences suitable for multi-modal calibration \& VI-SLAM evaluation. We also demonstrated new efficient algorithms for data synchronization during the acquisition process and a new iterative solution to estimate the unknown distortion coefficients of CARLA simulated cameras. Using three diverse weather conditions, we have shown their impact on ORB-SLAM3 and BASALT trajectory estimations.

The performance analysis includes a description of the sequence upon which the evaluation is done and the special conditions and corner cases simulated within every sequence to push the limits of the SLAM systems under assessment. The analytical study includes a comprehensive evaluation of the SLAM system performance and a qualitative comparison based on the ATE and RPE values. We hope that this new dataset will help in advancing the research in the field of multi-modal heterogeneous sensors fusion applied to Autonomous Ground Vehicles (AGVs) navigation in large-scale and dynamic environments. 

As a future research trend, it will be indispensable to develop new efficient multi-modal: calibration and SLAM algorithms based on the fusion of heterogeneous sensors with different caption and spectral technologies. That allows the SLAM system to better estimate the trajectory based on a reliable continuous-time 3D scene mapping. Finally, an in-depth investigation is needed concerning the effect of map loss on SLAM systems estimations during long-term navigation in large-scale and dynamic weather environments.

\backmatter
\bmhead{Supplementary information\label{secApndx1}} The open source data acquisition APIs and all sequences can be accessed using the Github repository: \url{https://github.com/AbanobSoliman/IBISCape.git}

In the repository there is a complete manual on how to execute the APIs in all setups and options, including the dataset files format and hierarchy.

\section*{Statements and Declarations}
\begin{itemize}

\item \textbf{Funding}
This work is supported by the French Ministry of Higher Education, Research and Innovation (MESRI). Author A.S. has received a Ph. D. grant from MESRI covering this research.

\item \textbf{Competing Interests}
The authors have no relevant financial or non-financial interests to disclose.

\item \textbf{Authors' contributions}
All authors contributed to the study conception and design. The first draft of the manuscript was written by A.S. and all authors commented on previous versions of the manuscript. All authors read and approved the final manuscript.

\item \textbf{Ethics approval} 
The submitted work is original and not have been published elsewhere in any form or language.

\item \textbf{Ethics Declarations}
This research work is based on computer simulation open source software and did not involve human participants or animals. Hence, Consent to participate and Consent for publication are not applicable.

\item \textbf{Consent to participate}
Not applicable.

\item \textbf{Consent for publication}
Not applicable.

\end{itemize}


\begin{thebibliography}{10}
\expandafter\ifx\csname url\endcsname\relax
  \def\url#1{\burl{#1}}\fi
\expandafter\ifx\csname urlprefix\endcsname\relax\def\urlprefix{URL }\fi
\providecommand{\bibinfo}[2]{#2}
\providecommand{\eprint}[2][]{\url{#2}}
\providecommand{\doi}[1]{\url{https://doi.org/#1}}
\bibcommenthead

\bibitem{7782863}
\bibinfo{author}{Forster, C.}, \bibinfo{author}{Zhang, Z.},
  \bibinfo{author}{Gassner, M.}, \bibinfo{author}{Werlberger, M.} \&
  \bibinfo{author}{Scaramuzza, D.}
\newblock \bibinfo{title}{Svo: Semidirect visual odometry for monocular and
  multicamera systems}.
\newblock \emph{\bibinfo{journal}{IEEE Transactions on Robotics}}
  \textbf{\bibinfo{volume}{33}}~(2), \bibinfo{pages}{249--265}
  (\bibinfo{year}{2017}).
\newblock \doi{10.1109/TRO.2016.2623335} .

\bibitem{Leutenegger2014}
\bibinfo{author}{Leutenegger, S.}, \bibinfo{author}{Lynen, S.},
  \bibinfo{author}{Bosse, M.}, \bibinfo{author}{Siegwart, R.} \&
  \bibinfo{author}{Furgale, P.}
\newblock \bibinfo{title}{Keyframe-based visual-inertial odometry using
  nonlinear optimization}.
\newblock \emph{\bibinfo{journal}{The International Journal of Robotics
  Research}} \textbf{\bibinfo{volume}{34}} (\bibinfo{year}{2014}).
\newblock \doi{10.1177/0278364914554813} .

\bibitem{7557075}
\bibinfo{author}{{Forster}, C.}, \bibinfo{author}{{Carlone}, L.},
  \bibinfo{author}{{Dellaert}, F.} \& \bibinfo{author}{{Scaramuzza}, D.}
\newblock \bibinfo{title}{On-manifold preintegration for real-time
  visual--inertial odometry}.
\newblock \emph{\bibinfo{journal}{IEEE Transactions on Robotics}}
  \textbf{\bibinfo{volume}{33}}~(1), \bibinfo{pages}{1--21}
  (\bibinfo{year}{2017}).
\newblock \doi{10.1109/TRO.2016.2597321} .

\bibitem{8421746}
\bibinfo{author}{{Qin}, T.}, \bibinfo{author}{{Li}, P.} \&
  \bibinfo{author}{{Shen}, S.}
\newblock \bibinfo{title}{Vins-mono: A robust and versatile monocular
  visual-inertial state estimator}.
\newblock \emph{\bibinfo{journal}{IEEE Transactions on Robotics}}
  \textbf{\bibinfo{volume}{34}}~(4), \bibinfo{pages}{1004--1020}
  (\bibinfo{year}{2018}).
\newblock \doi{10.1109/TRO.2018.2853729} .

\bibitem{Kerl2013}
\bibinfo{author}{{Kerl}, C.}, \bibinfo{author}{{Sturm}, J.} \&
  \bibinfo{author}{{Cremers}, D.}
\newblock \bibinfo{title}{Dense visual slam for rgb-d cameras}
  (\bibinfo{year}{2013}).
\newblock \bibinfo{note}{2013 IEEE/RSJ International Conference on Intelligent
  Robots and Systems}.

\bibitem{Alliez2020RealTimeMS}
\bibinfo{author}{Alliez, P.} \emph{et~al.}
\newblock \bibinfo{title}{Real-time multi-slam system for agent localization
  and 3d mapping in dynamic scenarios}.
\newblock \emph{\bibinfo{journal}{2020 IEEE/RSJ International Conference on
  Intelligent Robots and Systems (IROS)}} \bibinfo{pages}{4894--4900}
  (\bibinfo{year}{2020}) .

\bibitem{caron2006gps}
\bibinfo{author}{Caron, F.}, \bibinfo{author}{Duflos, E.},
  \bibinfo{author}{Pomorski, D.} \& \bibinfo{author}{Vanheeghe, P.}
\newblock \bibinfo{title}{Gps/imu data fusion using multisensor kalman
  filtering: introduction of contextual aspects}.
\newblock \emph{\bibinfo{journal}{Information fusion}}
  \textbf{\bibinfo{volume}{7}}~(2), \bibinfo{pages}{221--230}
  (\bibinfo{year}{2006}) .

\bibitem{yangicalib}
\bibinfo{author}{Yang, Y.} \emph{et~al.}
\newblock \bibinfo{title}{icalib: Inertial aided multi-sensor calibration}
  (\bibinfo{year}{2021}).
\newblock \bibinfo{note}{ICRA - VINS Workshop 2021, Xi'an, China}.

\bibitem{9387269}
\bibinfo{author}{Per\v{S}i\'c, J.}, \bibinfo{author}{Petrovi\'c, L.},
  \bibinfo{author}{Markovi\'c, I.} \& \bibinfo{author}{Petrovi\'c, I.}
\newblock \bibinfo{title}{Spatiotemporal multisensor calibration via gaussian
  processes moving target tracking}.
\newblock \emph{\bibinfo{journal}{IEEE Transactions on Robotics}}
  \bibinfo{pages}{1--15} (\bibinfo{year}{2021}).
\newblock \doi{10.1109/TRO.2021.3061364} .

\bibitem{leeefficient}
\bibinfo{author}{Lee, W.}, \bibinfo{author}{Yang, Y.} \&
  \bibinfo{author}{Huang, G.}
\newblock \bibinfo{title}{Efficient multi-sensor aided inertial navigation with
  online calibration} (\bibinfo{year}{2021}).
\newblock \bibinfo{note}{2021 IEEE International Conference on Robotics and
  Automation (ICRA)}.

\bibitem{gehrig2021combining}
\bibinfo{author}{Gehrig, D.}, \bibinfo{author}{R{\"u}egg, M.},
  \bibinfo{author}{Gehrig, M.}, \bibinfo{author}{Hidalgo-Carri{\'o}, J.} \&
  \bibinfo{author}{Scaramuzza, D.}
\newblock \bibinfo{title}{Combining events and frames using recurrent
  asynchronous multimodal networks for monocular depth prediction}.
\newblock \emph{\bibinfo{journal}{IEEE Robotics and Automation Letters}}
  \textbf{\bibinfo{volume}{6}}~(2), \bibinfo{pages}{2822--2829}
  (\bibinfo{year}{2021}) .

\bibitem{Gehrig21ral}
\bibinfo{author}{Gehrig, M.}, \bibinfo{author}{Aarents, W.},
  \bibinfo{author}{Gehrig, D.} \& \bibinfo{author}{Scaramuzza, D.}
\newblock \bibinfo{title}{Dsec: A stereo event camera dataset for driving
  scenarios}.
\newblock \emph{\bibinfo{journal}{IEEE Robotics and Automation Letters}}
  \textbf{\bibinfo{volume}{PP}}, \bibinfo{pages}{1--8} (\bibinfo{year}{2021}).
\newblock \doi{10.1109/LRA.2021.3068942} .

\bibitem{Li2021PlanarSLAM}
\bibinfo{author}{Li, Y.}, \bibinfo{author}{Yunus, R.}, \bibinfo{author}{Brasch,
  N.}, \bibinfo{author}{Navab, N.} \& \bibinfo{author}{Tombari, F.}
\newblock \bibinfo{title}{Rgb-d slam with structural regularities}.
\newblock \emph{\bibinfo{journal}{2021 IEEE International Conference on
  Robotics and Automation (ICRA)}} \bibinfo{pages}{11581--11587}
  (\bibinfo{year}{2021}) .

\bibitem{s20072068}
\bibinfo{author}{Debeunne, C.} \& \bibinfo{author}{Vivet, D.}
\newblock \bibinfo{title}{A review of visual-lidar fusion based simultaneous
  localization and mapping}.
\newblock \emph{\bibinfo{journal}{Sensors}} \textbf{\bibinfo{volume}{20}}~(7)
  (\bibinfo{year}{2020}).
\newblock \urlprefix\url{https://www.mdpi.com/1424-8220/20/7/2068}.
\newblock \doi{10.3390/s20072068} .

\bibitem{Dosovitskiy17}
\bibinfo{author}{Dosovitskiy, A.}, \bibinfo{author}{Ros, G.},
  \bibinfo{author}{Codevilla, F.}, \bibinfo{author}{Lopez, A.} \&
  \bibinfo{author}{Koltun, V.}
\newblock \bibinfo{title}{{CARLA}: {An} open urban driving simulator}
  (\bibinfo{year}{2017}).
\newblock \bibinfo{note}{Proceedings of the 1st Annual Conference on Robot
  Learning}.

\bibitem{sturm2012benchmark}
\bibinfo{author}{Sturm, J.}, \bibinfo{author}{Engelhard, N.},
  \bibinfo{author}{Endres, F.}, \bibinfo{author}{Burgard, W.} \&
  \bibinfo{author}{Cremers, D.}
\newblock \bibinfo{title}{A benchmark for the evaluation of rgb-d slam systems}
  (\bibinfo{year}{2012}).
\newblock \bibinfo{note}{2012 IEEE/RSJ international conference on intelligent
  robots and systems}.

\bibitem{geiger2013vision}
\bibinfo{author}{Geiger, A.}, \bibinfo{author}{Lenz, P.},
  \bibinfo{author}{Stiller, C.} \& \bibinfo{author}{Urtasun, R.}
\newblock \bibinfo{title}{Vision meets robotics: The kitti dataset}.
\newblock \emph{\bibinfo{journal}{The International Journal of Robotics
  Research}} \textbf{\bibinfo{volume}{32}}~(11), \bibinfo{pages}{1231--1237}
  (\bibinfo{year}{2013}) .

\bibitem{blanco2014malaga}
\bibinfo{author}{Blanco-Claraco, J.-L.}, \bibinfo{author}{Moreno-Duenas, F.-A.}
  \& \bibinfo{author}{Gonz{\'a}lez-Jim{\'e}nez, J.}
\newblock \bibinfo{title}{The m{\'a}laga urban dataset: High-rate stereo and
  lidar in a realistic urban scenario}.
\newblock \emph{\bibinfo{journal}{The International Journal of Robotics
  Research}} \textbf{\bibinfo{volume}{33}}~(2), \bibinfo{pages}{207--214}
  (\bibinfo{year}{2014}) .

\bibitem{carlevaris2016university}
\bibinfo{author}{Carlevaris-Bianco, N.}, \bibinfo{author}{Ushani, A.~K.} \&
  \bibinfo{author}{Eustice, R.~M.}
\newblock \bibinfo{title}{University of michigan north campus long-term vision
  and lidar dataset}.
\newblock \emph{\bibinfo{journal}{The International Journal of Robotics
  Research}} \textbf{\bibinfo{volume}{35}}~(9), \bibinfo{pages}{1023--1035}
  (\bibinfo{year}{2016}) .

\bibitem{burri2016euroc}
\bibinfo{author}{Burri, M.} \emph{et~al.}
\newblock \bibinfo{title}{The euroc micro aerial vehicle datasets}.
\newblock \emph{\bibinfo{journal}{The International Journal of Robotics
  Research}} \textbf{\bibinfo{volume}{35}}~(10), \bibinfo{pages}{1157--1163}
  (\bibinfo{year}{2016}) .

\bibitem{majdik2017zurich}
\bibinfo{author}{Majdik, A.~L.}, \bibinfo{author}{Till, C.} \&
  \bibinfo{author}{Scaramuzza, D.}
\newblock \bibinfo{title}{The zurich urban micro aerial vehicle dataset}.
\newblock \emph{\bibinfo{journal}{The International Journal of Robotics
  Research}} \textbf{\bibinfo{volume}{36}}~(3), \bibinfo{pages}{269--273}
  (\bibinfo{year}{2017}) .

\bibitem{pfrommer2017penncosyvio}
\bibinfo{author}{Pfrommer, B.}, \bibinfo{author}{Sanket, N.},
  \bibinfo{author}{Daniilidis, K.} \& \bibinfo{author}{Cleveland, J.}
\newblock \bibinfo{title}{Penncosyvio: A challenging visual inertial odometry
  benchmark} (\bibinfo{year}{2017}).
\newblock \bibinfo{note}{2017 IEEE International Conference on Robotics and
  Automation (ICRA)}.

\bibitem{schubert2018vidataset}
\bibinfo{author}{Schubert, D.} \emph{et~al.}
\newblock \bibinfo{title}{The tum vi benchmark for evaluating visual-inertial
  odometry}.
\newblock \emph{\bibinfo{journal}{2018 IEEE/RSJ International Conference on
  Intelligent Robots and Systems (IROS)}} \bibinfo{pages}{1680--1687}
  (\bibinfo{year}{2018}) .

\bibitem{judd2019oxford}
\bibinfo{author}{Judd, K.~M.} \& \bibinfo{author}{Gammell, J.~D.}
\newblock \bibinfo{title}{The oxford multimotion dataset: Multiple se (3)
  motions with ground truth}.
\newblock \emph{\bibinfo{journal}{IEEE Robotics and Automation Letters}}
  \textbf{\bibinfo{volume}{4}}~(2), \bibinfo{pages}{800--807}
  (\bibinfo{year}{2019}) .

\bibitem{jeong2019complex}
\bibinfo{author}{Jeong, J.}, \bibinfo{author}{Cho, Y.}, \bibinfo{author}{Shin,
  Y.-S.}, \bibinfo{author}{Roh, H.} \& \bibinfo{author}{Kim, A.}
\newblock \bibinfo{title}{Complex urban dataset with multi-level sensors from
  highly diverse urban environments}.
\newblock \emph{\bibinfo{journal}{The International Journal of Robotics
  Research}} \textbf{\bibinfo{volume}{38}}~(6), \bibinfo{pages}{642--657}
  (\bibinfo{year}{2019}) .

\bibitem{kasper2019benchmark}
\bibinfo{author}{Kasper, M.}, \bibinfo{author}{McGuire, S.} \&
  \bibinfo{author}{Heckman, C.}
\newblock \bibinfo{title}{A benchmark for visual-inertial odometry systems
  employing onboard illumination} (\bibinfo{year}{2019}).
\newblock \bibinfo{note}{2019 IEEE/RSJ International Conference on Intelligent
  Robots and Systems (IROS)}.

\bibitem{delmerico2019we}
\bibinfo{author}{Delmerico, J.}, \bibinfo{author}{Cieslewski, T.},
  \bibinfo{author}{Rebecq, H.}, \bibinfo{author}{Faessler, M.} \&
  \bibinfo{author}{Scaramuzza, D.}
\newblock \bibinfo{title}{Are we ready for autonomous drone racing? the uzh-fpv
  drone racing dataset} (\bibinfo{year}{2019}).
\newblock \bibinfo{note}{2019 International Conference on Robotics and
  Automation (ICRA)}.

\bibitem{zuniga2020vi}
\bibinfo{author}{Zu{\~n}iga-No{\"e}l, D.}, \bibinfo{author}{Jaenal, A.},
  \bibinfo{author}{Gomez-Ojeda, R.} \& \bibinfo{author}{Gonzalez-Jimenez, J.}
\newblock \bibinfo{title}{The uma-vi dataset: Visual--inertial odometry in
  low-textured and dynamic illumination environments}.
\newblock \emph{\bibinfo{journal}{The International Journal of Robotics
  Research}} \textbf{\bibinfo{volume}{39}}~(9), \bibinfo{pages}{1052--1060}
  (\bibinfo{year}{2020}) .

\bibitem{antonini2020blackbird}
\bibinfo{author}{Antonini, A.}, \bibinfo{author}{Guerra, W.},
  \bibinfo{author}{Murali, V.}, \bibinfo{author}{Sayre-McCord, T.} \&
  \bibinfo{author}{Karaman, S.}
\newblock \bibinfo{title}{The blackbird uav dataset}.
\newblock \emph{\bibinfo{journal}{The International Journal of Robotics
  Research}} \textbf{\bibinfo{volume}{39}}~(10-11), \bibinfo{pages}{1346--1364}
  (\bibinfo{year}{2020}) .

\bibitem{9341713}
\bibinfo{author}{Zhang, H.}, \bibinfo{author}{Jin, L.} \& \bibinfo{author}{Ye,
  C.}
\newblock \bibinfo{title}{The vcu-rvi benchmark: Evaluating visual inertial
  odometry for indoor navigation applications with an rgb-d camera}.
\newblock \emph{\bibinfo{journal}{2020 IEEE/RSJ International Conference on
  Intelligent Robots and Systems (IROS)}} \bibinfo{pages}{6209--6214}
  (\bibinfo{year}{2020}).
\newblock \doi{10.1109/IROS45743.2020.9341713} .

\bibitem{9351597}
\bibinfo{author}{Minoda, K.}, \bibinfo{author}{Schilling, F.},
  \bibinfo{author}{Wüest, V.}, \bibinfo{author}{Floreano, D.} \&
  \bibinfo{author}{Yairi, T.}
\newblock \bibinfo{title}{Viode: A simulated dataset to address the challenges
  of visual-inertial odometry in dynamic environments}.
\newblock \emph{\bibinfo{journal}{IEEE Robotics and Automation Letters}}
  \textbf{\bibinfo{volume}{6}}~(2), \bibinfo{pages}{1343--1350}
  (\bibinfo{year}{2021}).
\newblock \doi{10.1109/LRA.2021.3058073} .

\bibitem{klenk2021tumvie}
\bibinfo{author}{Klenk, S.}, \bibinfo{author}{Chui, J.},
  \bibinfo{author}{Demmel, N.} \& \bibinfo{author}{Cremers, D.}
\newblock \bibinfo{title}{Tum-vie: The tum stereo visual-inertial event
  dataset}.
\newblock \emph{\bibinfo{journal}{2021 IEEE/RSJ International Conference on
  Intelligent Robots and Systems (IROS)}} \bibinfo{pages}{8601--8608}
  (\bibinfo{year}{2021}).
\newblock \doi{10.1109/IROS51168.2021.9636728} .

\bibitem{articlecompare}
\bibinfo{author}{Yuan, C.} \emph{et~al.}
\newblock \bibinfo{title}{A novel fault-tolerant navigation and positioning
  method with stereo-camera/micro electro mechanical systems inertial
  measurement unit (mems-imu) in hostile environment}.
\newblock \emph{\bibinfo{journal}{Micromachines}} \textbf{\bibinfo{volume}{9}},
  \bibinfo{pages}{626} (\bibinfo{year}{2018}).
\newblock \doi{10.3390/mi9120626} .

\bibitem{autonomousproposed}
\bibinfo{author}{Faessler, M.} \emph{et~al.}
\newblock \bibinfo{title}{Autonomous, vision-based flight and live dense 3d
  mapping with a quadrotor micro aerial vehicle}.
\newblock \emph{\bibinfo{journal}{Journal of Field Robotics}}
  \textbf{\bibinfo{volume}{33}}~(4), \bibinfo{pages}{431--450}
  (\bibinfo{year}{2016}).
\newblock
  \urlprefix\url{https://onlinelibrary.wiley.com/doi/abs/10.1002/rob.21581}.
\newblock \doi{rob.21581} .

\bibitem{6696917}
\bibinfo{author}{Lynen, S.}, \bibinfo{author}{Achtelik, M.~W.},
  \bibinfo{author}{Weiss, S.}, \bibinfo{author}{Chli, M.} \&
  \bibinfo{author}{Siegwart, R.}
\newblock \bibinfo{title}{A robust and modular multi-sensor fusion approach
  applied to mav navigation} (\bibinfo{year}{2013}).
\newblock \bibinfo{note}{2013 IEEE/RSJ International Conference on Intelligent
  Robots and Systems}.

\bibitem{mourikis2007multi}
\bibinfo{author}{Mourikis, A.~I.} \& \bibinfo{author}{Roumeliotis, S.~I.}
\newblock \bibinfo{title}{A multi-state constraint kalman filter for
  vision-aided inertial navigation} (\bibinfo{year}{2007}).
\newblock \bibinfo{note}{Proceedings 2007 IEEE International Conference on
  Robotics and Automation}.

\bibitem{bloesch2015robust}
\bibinfo{author}{Bloesch, M.}, \bibinfo{author}{Omari, S.},
  \bibinfo{author}{Hutter, M.} \& \bibinfo{author}{Siegwart, R.}
\newblock \bibinfo{title}{Robust visual inertial odometry using a direct
  ekf-based approach} (\bibinfo{year}{2015}).
\newblock \bibinfo{note}{2015 IEEE/RSJ international conference on intelligent
  robots and systems (IROS)}.

\bibitem{qin2018vins}
\bibinfo{author}{Qin, T.}, \bibinfo{author}{Li, P.} \& \bibinfo{author}{Shen,
  S.}
\newblock \bibinfo{title}{Vins-mono: A robust and versatile monocular
  visual-inertial state estimator}.
\newblock \emph{\bibinfo{journal}{IEEE Transactions on Robotics}}
  \textbf{\bibinfo{volume}{34}}~(4), \bibinfo{pages}{1004--1020}
  (\bibinfo{year}{2018}) .

\bibitem{leutenegger2015keyframe}
\bibinfo{author}{Leutenegger, S.}, \bibinfo{author}{Lynen, S.},
  \bibinfo{author}{Bosse, M.}, \bibinfo{author}{Siegwart, R.} \&
  \bibinfo{author}{Furgale, P.}
\newblock \bibinfo{title}{Keyframe-based visual--inertial odometry using
  nonlinear optimization}.
\newblock \emph{\bibinfo{journal}{The International Journal of Robotics
  Research}} \textbf{\bibinfo{volume}{34}}~(3), \bibinfo{pages}{314--334}
  (\bibinfo{year}{2015}) .

\bibitem{Campos_2021}
\bibinfo{author}{Campos, C.}, \bibinfo{author}{Elvira, R.},
  \bibinfo{author}{Rodriguez, J. J.~G.}, \bibinfo{author}{M.~Montiel, J.~M.} \&
  \bibinfo{author}{D.~Tardos, J.}
\newblock \bibinfo{title}{Orb-slam3: An accurate open-source library for
  visual, visual-inertial, and multimap slam}.
\newblock \emph{\bibinfo{journal}{IEEE Transactions on Robotics}}
  \bibinfo{pages}{1--17} (\bibinfo{year}{2021}).
\newblock \urlprefix\url{http://dx.doi.org/10.1109/TRO.2021.3075644}.
\newblock \doi{10.1109/tro.2021.3075644} .

\bibitem{usenko19nfr}
\bibinfo{author}{Usenko, V.}, \bibinfo{author}{Demmel, N.},
  \bibinfo{author}{Schubert, D.}, \bibinfo{author}{Stueckler, J.} \&
  \bibinfo{author}{Cremers, D.}
\newblock \bibinfo{title}{Visual-inertial mapping with non-linear factor
  recovery}.
\newblock \emph{\bibinfo{journal}{IEEE Robotics and Automation Letters (RA-L)
  \& Int. Conference on Intelligent Robotics and Automation (ICRA)}}
  \textbf{\bibinfo{volume}{5}}~(2), \bibinfo{pages}{422--429}
  (\bibinfo{year}{2020}).
\newblock \doi{10.1109/LRA.2019.2961227} .

\bibitem{8460664}
\bibinfo{author}{Delmerico, J.} \& \bibinfo{author}{Scaramuzza, D.}
\newblock \bibinfo{title}{A benchmark comparison of monocular visual-inertial
  odometry algorithms for flying robots} (\bibinfo{year}{2018}).
\newblock \bibinfo{note}{2018 IEEE International Conference on Robotics and
  Automation (ICRA)}.

\bibitem{Zhou21tro}
\bibinfo{author}{Zhou, Y.}, \bibinfo{author}{Gallego, G.} \&
  \bibinfo{author}{Shen, S.}
\newblock \bibinfo{title}{Event-based stereo visual odometry}.
\newblock \emph{\bibinfo{journal}{IEEE Transactions on Robotics}}
  \textbf{\bibinfo{volume}{37}}~(5), \bibinfo{pages}{1433--1450}
  (\bibinfo{year}{2021}).
\newblock \doi{10.1109/TRO.2021.3062252} .

\bibitem{Gehrig_2020_CVPR}
\bibinfo{author}{Gehrig, D.}, \bibinfo{author}{Gehrig, M.},
  \bibinfo{author}{Hidalgo-Carrio, J.} \& \bibinfo{author}{Scaramuzza, D.}
\newblock \bibinfo{title}{Video to events: Recycling video datasets for event
  cameras} (\bibinfo{year}{2020}).
\newblock \bibinfo{note}{{IEEE} Conf. Comput. Vis. Pattern Recog. (CVPR)}.

\bibitem{rehder2016extending}
\bibinfo{author}{Rehder, J.}, \bibinfo{author}{Nikolic, J.},
  \bibinfo{author}{Schneider, T.}, \bibinfo{author}{Hinzmann, T.} \&
  \bibinfo{author}{Siegwart, R.}
\newblock \bibinfo{title}{Extending kalibr: Calibrating the extrinsics of
  multiple imus and of individual axes} (\bibinfo{year}{2016}).
\newblock \bibinfo{note}{2016 IEEE International Conference on Robotics and
  Automation (ICRA)}.

\bibitem{Muglikar2021CVPR}
\bibinfo{author}{Muglikar, M.}, \bibinfo{author}{Gehrig, M.},
  \bibinfo{author}{Gehrig, D.} \& \bibinfo{author}{Scaramuzza, D.}
\newblock \bibinfo{title}{How to calibrate your event camera}.
\newblock \emph{\bibinfo{journal}{2021 IEEE/CVF Conference on Computer Vision
  and Pattern Recognition Workshops (CVPRW)}} \bibinfo{pages}{1403--1409}
  (\bibinfo{year}{2021}) .

\bibitem{Rebecq19cvpr}
\bibinfo{author}{Rebecq, H.}, \bibinfo{author}{Ranftl, R.},
  \bibinfo{author}{Koltun, V.} \& \bibinfo{author}{Scaramuzza, D.}
\newblock \bibinfo{title}{Events-to-video: Bringing modern computer vision to
  event cameras}.
\newblock \emph{\bibinfo{journal}{{IEEE} Conf. Comput. Vis. Pattern Recog.
  (CVPR)}}  (\bibinfo{year}{2019}) .

\end{thebibliography}
\end{document}